\documentclass[letterpaper]{article} % DO NOT CHANGE THIS
\usepackage{aaai2026}  % DO NOT CHANGE THIS
\usepackage{times}  % DO NOT CHANGE THIS
\usepackage{helvet}  % DO NOT CHANGE THIS
\usepackage{courier}  % DO NOT CHANGE THIS
\usepackage[hyphens]{url}  % DO NOT CHANGE THIS
\usepackage{graphicx} % DO NOT CHANGE THIS
\usepackage{natbib}  % DO NOT CHANGE THIS AND DO NOT ADD ANY OPTIONS TO IT
\usepackage{caption} % DO NOT CHANGE THIS AND DO NOT ADD ANY OPTIONS TO IT
\usepackage{algorithm}
\usepackage{algorithmic}
\usepackage{amsmath}
\usepackage{amsfonts}
\usepackage{multirow}
\usepackage{graphicx}

\usepackage{amsmath}
\usepackage{url}
\usepackage{colortbl}  % 需要 colortbl 宏包
\usepackage{xcolor}    % 提供颜色扩展
\usepackage{subcaption}  % 推荐使用这个包，格式更现代

\usepackage{listings}
\usepackage{booktabs}
\usepackage{float}

\usepackage{newfloat}
\usepackage{listings}
\DeclareCaptionStyle{ruled}{labelfont=normalfont,labelsep=colon,strut=off} % DO NOT CHANGE THIS
\floatstyle{ruled}
\newfloat{listing}{tb}{lst}{}
\floatname{listing}{Listing}
\newcounter{corrfn}
\makeatletter  
\def\corresponding{%
  \ifnum\value{corrfn}=0% 
    \footnote{They are co-corresponding authors}%
    \setcounter{corrfn}{\value{footnote}}% 记录当前脚注编号
  \else% 后续调用：仅引用已生成的脚注编号
    \footnotemark[\value{corrfn}]%
  \fi%
}
\makeatother 

\title{HogVul: {Black-box} Adversarial Code {Generation Framework} Against LM-based Vulnerability Detectors}
\author{
    Jingxiao Yang\textsuperscript{\rm 1}\equalcontrib, 
    Ping He\textsuperscript{\rm 1}\equalcontrib, 
    Tianyu Du\textsuperscript{\rm 1,3}\corresponding, 
    Sun Bing\textsuperscript{\rm 2}, 
    Xuhong Zhang\textsuperscript{\rm 1,3}\corresponding 
}
\affiliations{
    \textsuperscript{\rm 1}Zhejiang University\\
    \textsuperscript{\rm 2}National Certification Technology (Hangzhou) Co., Ltd\\
    \textsuperscript{\rm 3}Ningbo Global Innovation Center, Zhejiang University\\
    \{zjradty, zhangxuhong\}@zju.edu.cn

}

\usepackage{bibentry}
\begin{document}

\maketitle

\begin{abstract}
Recent advances in software vulnerability detection have been driven by Language Model (LM)-based approaches.
However, these models remain vulnerable to adversarial attacks that exploit lexical and syntax perturbations, allowing critical flaws to evade detection.
Existing black-box attacks on LM-based vulnerability detectors primarily rely on isolated perturbation strategies, limiting their ability to efficiently explore the adversarial code space for optimal perturbations.
To bridge this gap, we propose HogVul, a black-box adversarial code generation framework that integrates both lexical and syntax perturbations under a unified dual-channel optimization strategy driven by Particle Swarm Optimization (PSO).
By systematically coordinating two-level perturbations, HogVul effectively expands the search space for adversarial examples, enhancing the attack efficacy.
Extensive experiments on four benchmark datasets demonstrate that HogVul achieves an average attack success rate improvement of 26.05\% over state-of-the-art baseline methods.
These findings highlight the potential of hybrid optimization strategies in exposing model vulnerabilities.
\end{abstract}

\section{Introduction}
\label{sec:intro}

\begin{figure}[t] 
\centering
  \includegraphics[width=\columnwidth, height=\textheight, keepaspectratio]{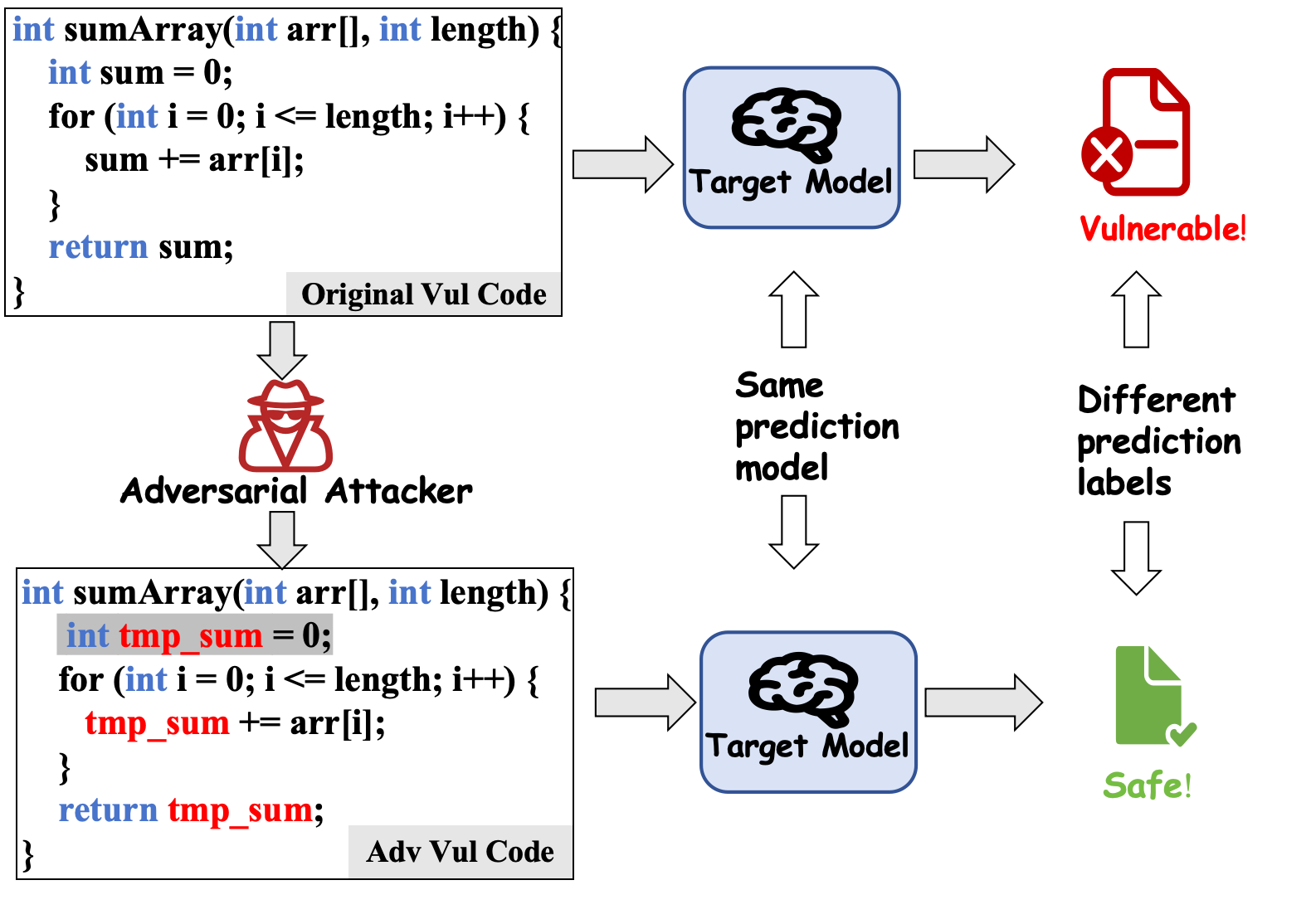}
  \caption{A simple example of adversarial code misleading models.}
  \label{fig:case}
\end{figure}

Software vulnerability detection has become increasingly critical with the surge in security breaches~\citep{thestackcew2023}. While language model (LM)-based approaches achieve state-of-the-art results in this domain~\cite{feng2020codebert, guo2020graphcodebert, wang2021codet5, steenhoek2023empiricalstudydeeplearning}, they remain susceptible to adversarial attacks that subtly modify code to bypass detection~\cite{introshayegani2023surveyvulnerabilitieslargelanguage, introchacko2024adversarialattackslargelanguage,Tencent_AI-Infra-Guard_2025,tan2025dyp,tan2025bottom}.
Figure~\ref{fig:case} illustrates how a simple array index out-of-bounds error can evade detection through minor code changes, leading to severe security risks.

In practice, LM-based detectors are typically black-box systems, making them vulnerable to targeted black-box attacks~\cite{mhm, alert, dip, codeattack, coda, codetae}.
However, these methods only achieve limited attack effectiveness, which is insufficient to provide a realistic evaluation for the following reasons.
The existing attack methods only consider a single limited perturbation.
For instance, ALERT~\cite{alert} only considers the lexical perturbation, which cannot provide the syntax-level evaluation.
Additionally, the existing methods fail to propose an efficient optimization for the optimal adversarial perturbation generation.
For instance, DIP~\cite{dip} relies primarily on randomly sampling and sequentially inserting dead code fragments at the syntax-level, which lacks an explicit guided search mechanism to locate effective code segments efficiently.

To overcome these limitations, we introduce \textbf{HogVul}, a black-box adversarial code generation framework that integrates lexical and syntax perturbations within a unified optimization process to effectively evaluate the robustness of LM-based vulnerability detectors.
Our design is motivated by the observation that language models are vulnerable at multiple levels, including lexical patterns and syntactic dependencies~\cite{naturalcc}.
Targeting only one of these levels often fails to sufficiently disrupt model behavior, especially in models pretrained on rich structural and semantic patterns~\cite{liu2024multigranularadversarialattacksblackbox}.
From the LM perspective, multi-level attacks are more likely to disrupt deeper attention layers and introduce representational uncertainty~\cite{kim2022diversegenerativeperturbationsattention}.
Regarding the perturbation perspective, this enables the fusion of heterogeneous perturbation signals, expanding the search space and increasing the likelihood of discovering high-quality adversarial solutions~\cite{mao2020compositeadversarialattacks}.
However, the expanded space also poses a convergence challenge for optimization.
To address this, we reformulate adversarial code search as an evolutionary optimization problem, where information is shared across particles to iteratively narrow the search scope and guide convergence toward compact regions of optimal perturbations. 
Specifically, HogVul expands perturbation to manipulate code features at both lexical and syntax levels, collectively challenging the model's classification boundaries.
To efficiently find the optimal solutions in the hybrid perturbation space, HogVul designs a customized hybrid optimization method based on particle swarm optimization (PSO)~\cite{pso1995}.
Our method integrates both perturbation types into a unified optimization loop, using a stagnation counter to monitor progress and trigger switches between perturbation strategies.
Once a globally optimal particle is identified, HogVul leverages particle level information sharing and stagnation triggered switching to avoid redundant exploration and guide the swarm towards more promising regions.

To rigorously validate the efficiency and quality of HogVul, we design a comprehensive evaluation consisting of five complementary aspects.
First, we assess the effectiveness of HogVul by conducting attacks across three widely-used benchmark datasets spanning diverse vulnerability types~\citep{mitre2023cwe}.
The results demonstrate that HogVul achieves an average Attack Success Rate (ASR) improvement of 26.05\% compared to baseline methods.
Second, we provide empirical evidence of PSO’s optimization rationality by visualizing search trajectories, showing that it effectively guides perturbations toward high-impact configurations.
Third, we conduct ablation studies to quantify the contribution of key components. Results reveal that each component play crucial yet complementary roles, as removing either component noticeably reduces overall ASR.
Fourth, we assess HogVul’s robustness under distribution shifts by applying it to a completely unseen dataset, where it maintains high ASR without parameter tuning, demonstrating strong robustness against diverse vulnerability patterns.
Finally, we simulate real-world deployment scenarios by measuring the increase in False Negative Rate (FNR) when HogVul is applied to vulnerable code, The results underscore its potential to evade detection systems by generating highly effective adversarial samples.
These experimental results jointly confirm HogVul’s robustness, effectiveness, and practical applicability under diverse attack settings and real-world conditions.

Our contributions are illustrated as follows.

\begin{itemize}
    \item We propose HogVul, a black-box adversarial code generation method for evaluating the robustness of LM-based vulnerability detectors.
    \item HogVul considers both lexical and syntax level, enabling a broader perturbation search space to discover more weaknesses of the LM-based vulnerability detectors.
    \item To find the optimal perturbations, we develop an efficient hybrid optimization framework that unifies both perturbation types, ensuring continuous and relevant exploration of the perturbation space.
\end{itemize}

\section{Related Work}
\label{sec:related}

\textbf{Black-box Adversarial Attacks on Language Models.} Recent black-box adversarial attacks on language models fall into two categories: lexical-level and syntax-level perturbations. Lexical methods modify tokens, while syntax-level approaches alter control flow or AST structures. However, existing methods often focus on one perturbation type or lack systematic optimization, motivating HogVul as a unified framework integrating both.

\textbf{Lexical-Level Perturbation.} These attacks use identifier renaming and token substitution to mislead models while preserving functionality. MHM~\cite{mhm} applies Metropolis-Hastings~\cite{metropolis1953equation} to iteratively rename identifiers via semantic similarity. ALERT~\cite{alert} generates candidates with pre-trained models and ranks them by cosine similarity using hybrid greedy-genetic search. Evolutionary methods~\cite{mercuri2023evolutionary} refine this with fitness functions balancing efficacy and semantics. RNNS~\cite{zhang2023blackboxattackcodemodels} builds a real-world code-based search space guided by Representation Nearest Neighbors. MOOA~\cite{zhou2024evolutionary} formulates lexical attacks as multi-objective optimization. Though effective at token-level semantics, they lack structural analysis, limiting performance in complex code~\cite{yefet2020adversarialexamplesmodelscode}.

\textbf{Syntax-Level Perturbation.} These perturbations disrupt code structure. AdVulCode~\cite{electronics12040936} uses Monte Carlo Tree Search for control-flow transformations. DIP~\cite{dip} inserts dead code via CodeBERT localization to maximize divergence. GraphCodeAttack~\cite{nguyen2024adversarialattackscodemodels} injects dead code at sensitive AST nodes using gradient-free influence analysis. Although syntax-level attacks can disrupt higher-level structural dependencies, they are often limited by the lack of coordinated optimization strategies, resulting in suboptimal perturbation generation.

Though both perturbation types have advanced, they are typically treated independently, ignoring synergies. HogVul bridges this gap with a PSO-driven framework integrating both, using stagnation-aware switching to dynamically alternate strategies. This dual-channel design expands the search space and boosts attack efficacy.

\section{Methodology}
\label{sec:methodology}

\begin{figure*}
     \centering
     \includegraphics[width=\textwidth, height=0.3\textheight, keepaspectratio]{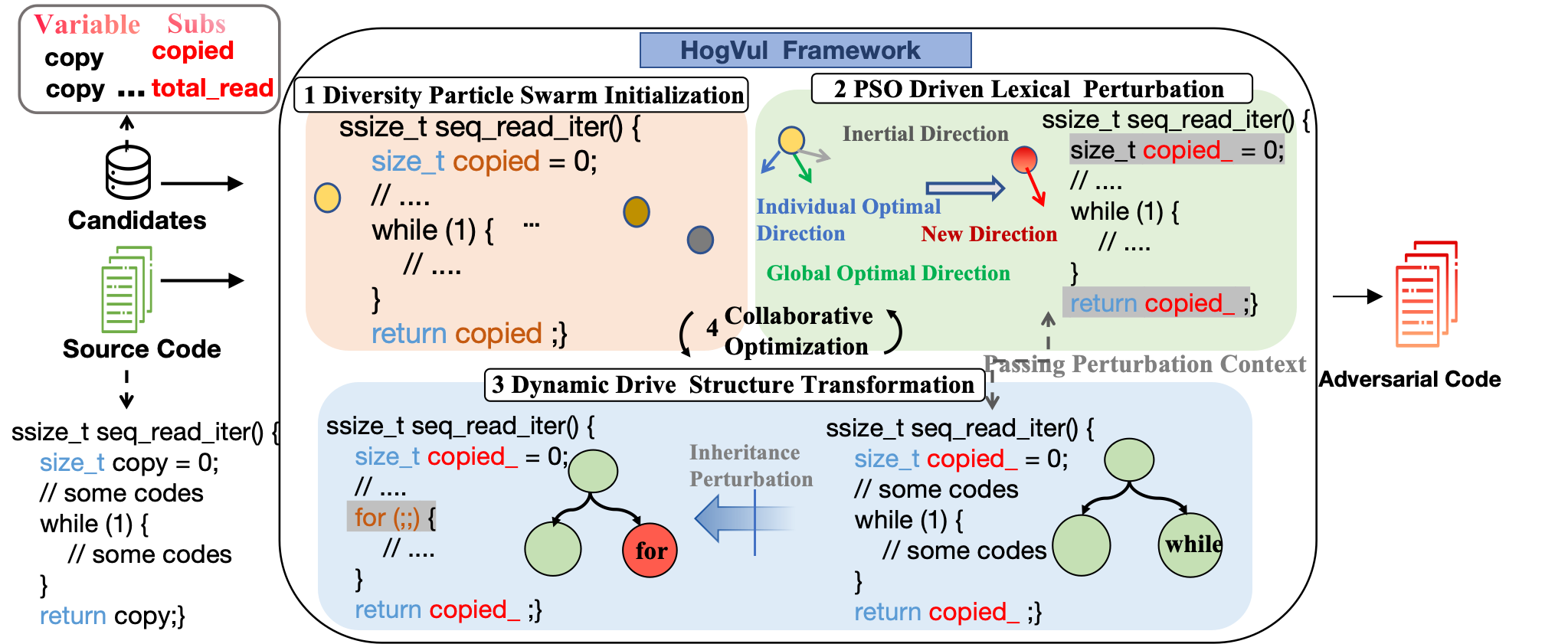}
    \caption{
     The overview of HogVul. It operates in four stages: \textcircled{1}Diversity-enhanced Initialization, \textcircled{2}PSO Driven Lexical Perturbation, \textcircled{3}Structure-Aware Code Transformation and \textcircled{4}Dual-Channel Cooperative Optimization. 
    }
    \label{fig:framework}
\end{figure*}

\subsection{Framework Overview}
\label{sec:problem formulation}

\textbf{Goal}. Our objective is to generate adversarial code examples that mislead the LM-based vulnerability detector $M$ into incorrect classification, while preserving the original program functionality and ensuring compilability. Specifically, given an input code sample $c$ with ground truth label $y \in \{0,1\}$, where $1$ denotes vulnerable and $0$ denotes secure. We aim to construct an adversarial code $c_{adv}$ such that:
\begin{equation} 
\begin{aligned} 
{argmax} \quad \mathbb{P}(M(c_{adv}) = 1 - y) \\ 
\text{s.t.} \quad c_{adv}=c+\delta,\quad \mathcal{F}(c_{adv}) = \mathcal{F}(c)
\end{aligned} 
\end{equation}
Where $M$ denotes the target model, $\mathbb{P}$ denotes the prediction probability, and $\mathcal{F}$ represents the program’s functionality. The perturbation $\delta$ represents the adversarial perturbation applied to the source code.

\noindent \textbf{Design}. The previous approach primarily focuses on a single perturbation to generate adversarial examples and lacks systematic optimization strategies, often relying on random or heuristic search.
To address these limitations, we propose \textbf{HogVul}, a query-based black-box attack framework that integrates multiple perturbation strategies under a unified dual-channel optimization process.
As shown in Figure~\ref{fig:framework}, HogVul proceeds through four stages. 
% (1) diversity-enhanced initialization, (2) lexical perturbation, (3) structural transformation and (4) cooperative optimization.
The lexical and structural channels operate in parallel, each maintaining its own candidate population while sharing a global best solution.
A stagnation-aware counter monitors progress in each channel, enabling dynamic switching to the complementary channel when local optima are detected.
This facilitates cross-channel cooperation and avoids premature convergence.

% 3.2
\subsection{Diversity-Promoting Initialization}
\label{sec:Diversity Particle Swarm Initialization}
Diverse initialization encourages exploration of multiple regions in the code search space, reducing the risk of the swarm getting trapped in local optima early in the optimization.
To achieve this, we propose a semantic-aware initialization strategy that exploits the sensitivity of code models to identifier semantics.
Prior work~\cite{allamanis2019adverseeffectscodeduplication} has shown that even minor changes to variable names can significantly alter transformer-based model predictions by shifting input embeddings and early attention.
Leveraging this property, we construct diverse initial populations by replacing variable names using a semantic similarity metric.
We utilize FastText embeddings~\cite{bojanowski2017enrichingwordvectorssubword} to encode each identifier $v$ as $\vec{v}$, and retrieve candidate substitutions $v'$ from a predefined vocabulary (following ALERT~\cite{alert}) based on cosine similarity.
Specifically, we compute:

\begin{equation}  
\text{Sim}(v, v') = 1 - \frac{\vec{v} \cdot \vec{v'}}{\|\vec{v}\| \cdot \|\vec{v'}\|}  
\end{equation}   

\noindent This similarity score lies in $[0, 2]$, where lower values indicate greater semantic closeness.

To balance exploitation of known high-quality substitutions with exploration of novel variants, we employ a two-fold substitution strategy: \textbf{(1) Similarity-guided initialization}. A subset of particles is initialized using top-$k$ most semantically similar substitutes for each identifier, ensuring high functional plausibility. \textbf{(2) Randomized sampling}. The remaining particles randomly sample replacements from the candidate pool to inject noise and increase lexical variance.
This design ensures broad coverage of the perturbation space of the code, providing a robust starting point for downstream optimization.

% 3.3
\subsection{PSO Driven Lexical Perturbation}
Generating adversarial examples in the code task is uniquely challenging due to the vast and combinatorial search space formed by potential lexical substitutions.
To effectively navigate this search space, we reformulate lexical perturbation as a population-based optimization problem, where the objective is to identify variable renaming strategies that maximize misclassification confidence.
While the section~\ref{sec:Diversity Particle Swarm Initialization} enhances the diversity in the initialization, efficiently exploring this large, discrete space still requires a principled approach.
Drawing on the success of PSO in adversarial natural language processing~\cite{psotext}, we adopt it as the core driver for refining lexical perturbation strategies.
Each particle represents a candidate adversarial code defined by variable renamings.
The process is driven by three components: \textbf{Inertia} preserves the particle’s previous direction in the search space; \textbf{Individual Best} guides the particle toward its most effective perturbation configuration; \textbf{Global Best} identifies the most impactful perturbation observed across the swarm, aligning with the attack objective by maximizing the model’s confidence drop.
During each iteration, the particle’s velocity is calculated as a weighted combination of the three components, which govern the particle’s update, specifying the adjustment direction for each lexical substitution.
For instance, if a particle previously substituted \texttt{copied} $\rightarrow$ \texttt{read\_bytes} but another particle’s best result applied \texttt{copied} $\rightarrow$ \texttt{total\_read}, the velocity update will adjust to increase the likelihood of adopting \texttt{total\_read}.
These three components jointly guide the evolution of renaming decisions across iterations.
These components determine the particle’s velocity, which specifies the adjustment direction for each lexical substitution, converging toward more adversarial configurations.
To enhance convergence, we apply adaptive tuning for PSO’s parameters $w, c_1, c_2$ based on swarm's progress, promoting wide exploration in early iterations and accelerates convergence as the swarm stabilizes.
A detailed description is available in Appendix A.
% ~\ref{sec:design of PSO}

% 3.4
\subsection{Structure-Aware Code Transformation}
Beyond lexical perturbation, we introduce a syntax-level perturbation that manipulates the control structures and control layout of source code, enriching the code search space by introducing syntactically different code variants. This can introduce larger shifts in model representations. Moreover, alternating between lexical and structural edits makes collaborative optimization possible: lexical attacks benefit from modified syntax contexts, while structural edits inherit effective variable renamings. We propose a four-step syntax perturbation that leverages both code syntax characteristics and model sensitivity:
\noindent\textbf{(1) Structural Type Extraction.}
We first parse the input code using TXL~\footnote{https://txl.ca/}. Guided by prior work~\citep{DBLP}, we extract control-relevant structures (e.g., if-else, for-loop) and compute their frequencies. This forms a structural profile for each program instance.
\noindent\textbf{(2) Transformation Probability Modeling.}
To prioritize impactful perturbations, we assign sampling probabilities to each structure type based on its frequency and importance. Control flow elements (as defined in Table~\ref{tab:control_flow_example}) are given extra weight, as they significantly influence program behavior and model predictions~\cite{naturalcc}. The sampling probability for structure $a_i$ is defined as:

\begin{equation}
\begin{aligned}
    P(a_i) &= \frac{\phi(s_i)}{\sum_{j=1}^{n} \phi(s_j)}, \\
    \text{where} \quad \phi(s_i) &=
    \begin{cases}
        \lambda \cdot s_i, & \text{if control-flow}, \\
        s_i, & \text{otherwise}.
    \end{cases}
\end{aligned}
\label{eq:probability-distribution}
\end{equation}
Here, $s_i$ denotes the occurrence count of structure $i$, and $\lambda=1.8$ adjusts the bias toward control flow elements. We analyze the impact of $\lambda$ in Appendix D.
~\ref{sec:ablation experiment}

\noindent\textbf{(3) Sensitive Location Identification.}
To localize high-impact perturbation sites, we apply a masking-based gradient analysis that reveals regions where structural changes can cause maximal confidence drop in the model. This ensures that transformations are applied where they are most adversarially potent. 
\noindent\textbf{(4) Targeted Transformation Execution.}
Based on the sampled structural type and the identified sensitive location, we apply a semantic-preserving transformation (e.g., loop restructuring). The full procedure is described in Algorithm~\ref{algo:structure-aware}.

% 3.5
\subsection{Dual-Channel Cooperative Optimization}
\label{sec:section 3.5}
Existing methods typically treat lexical and syntax perturbations independently, thereby missing the opportunity for cross-level information sharing and coordinated exploration. However, in transformer-based models, different types of perturbations affect distinct layers of attention and representation. Lexical substitutions influence the input token embeddings and early attention layers, while syntactic changes impact higher-level structural reasoning~\cite{alleman-etal-2021-syntactic,naturalcc}. Optimizing them in isolation risks redundancy or suboptimal exploration. To address this, we design a cooperative mechanism where two channels are optimized in the same perturbation space. Each channel maintains its own population of candidates but shares a global best solution, updated across both. This global solution serves as a bridge, allowing information to flow between the two search spaces. When either channel fails to improve for a fixed number of iterations, it adopts the best perturbation found by the other channel as a partial guidance, dynamically adjusting its search trajectory.

\noindent\textbf{Stagnation-Aware Channel Switching.} To monitor search progress, we track search progress using a stagnation counter:
\begin{equation}
\text{counter} =
\begin{cases} 
    \text{counter} + 1, & \text{if unchanged}, \\ 
    0, & \text{otherwise}.
\end{cases}
\label{eq:stagnation-counter}
\end{equation}
If no decline in model confidence occurs after a fixed number of iterations, the search adaptively switches to the complementary channel (e.g., from lexical to syntax), helping escape local optima and restoring exploration capacity.

\noindent\textbf{Cross-Channel Information Fusion.}
To preserve semantic continuity across perturbation levels, we enable bidirectional information flow between lexical and syntax channels. Rather than treating them as isolated stages, each channel leverages the other's outputs as contextual priors: when one channel stagnates, it inherits the best candidate from the other, including modified code, renamed identifiers, as a warm start for further exploration.
This design facilitates iterative refinement across perturbation levels. Lexical edits guide structural transformations with aligned semantics, while structural changes feed back into subsequent lexical updates with adapted context. By maintaining shared progress across channels, the search avoids semantic drift and resets, enabling more coherent and targeted search trajectories.

\section{Experiments}
\label{sec:experiments}
\subsection{Experimental Setups}
\textbf{Datasets and Target Models}:  
We evaluate our proposed HogVul on four widely used benchmark datasets:
\textbf{Devign}~\cite{devign}: A benchmark dataset consisting of over 48,000 functions collected from open-source C projects (FFmpeg and QEMU), widely used for evaluating vulnerability classification models.
    
\textbf{DiverseVul}~\cite{diversevul}: A large-scale and diverse C/C++ vulnerability dataset with over 18,000 vulnerable and 330,000 non-vulnerable functions across 797 projects, covering more than 150 CWE categories.
    
\textbf{BigVul}~\cite{bigvul}: An extensive dataset of C/C++ functions extracted from 348 real-world open-source projects, focusing on historical CVE-labeled vulnerabilities and widely used in vulnerability prediction tasks.
    
\textbf{D2A}~\cite{D2A}: A benchmark designed for differential vulnerability analysis, used to evaluate generalization ability on unseen code samples and assess robustness under real-world vulnerability settings.

To ensure consistency, we preprocess all datasets following the methodology in ALERT~\cite{alert} and adopt an 80\%, 10\%, 10\% split for training, validation and testing. These datasets contain a wide range of mutually exclusive real-world vulnerabilities. For more details, please refer to Appendix B.1.
% Appendix~\ref{sec:appendix_datasets}
 
We select CodeBERT~\cite{feng2020codebert}, GraphCodeBERT~\cite{guo2020graphcodebert}, and CodeT5~\cite{wang2021codet5} as target models due to their representativeness and wide adoption in vulnerability detection and other code intelligence tasks. They cover both encoder-only and encoder-decoder architectures, enabling a comprehensive evaluation across different model paradigms. All three are pretrained on large-scale code corpora, ensuring strong generalization and practical relevance.

\noindent\textbf{Baseline Methods}: We compare our proposed method with the following state-of-the-art adversarial attack techniques on code models, which are chosen to represent two complementary perturbation paradigms: 

\textbf{{ALERT}}~\cite{alert}: A lexical-level adversarial attack method that leverages pretrained language models to generate context-aware variable substitutions. It combines greedy and genetic algorithms to optimize attack effectiveness while preserving code naturalness and readability.

\textbf{DIP}~\cite{dip}: A syntax-level attack that inserts semantically neutral dead code to mislead vulnerability detectors. It ensures the generated adversarial examples remain compilable and syntactically valid.

The details for choosing the two baseline methods can be found in Appendix B.2.
% ~\ref{sec:appendix_baseline}

\noindent\textbf{Metrics}: We assess the effectiveness and efficiency of our attack framework using five evaluation metrics:  
(1) Attack Success Rate (ASR\%) ,  
(2) Average Confidence Drop ($\Delta_{\text{drop}}$),  
(3) CodeBLEU~\cite{CodeBLEU},  
(4) Code Average Diversity (CAD), and  
(5) ASR per Query (APQ).  
Notably, APQ and CAD are newly introduced metrics in this work to quantify query efficiency and diversity in adversarial samples, respectively. APQ captures the trade-off between attack success and query cost, enabling fair comparisons across methods with varying query budgets or optimization strategies. CAD quantifies the population-level diversity of adversarial examples.
Detailed definitions for all metrics are provided in Appendix B.3.
% ~\ref{sec:mertics}

All experiments are conducted on a server equipped with an NVIDIA A800 GPU. For each attack method, we perform ten independent executions on each model and report the average results to ensure robustness and reliability.

\section{Analysis}
\label{sec:analysis}

\begin{table*}[ht]
    \small
    \centering
    \begin{tabular*}{\textwidth}{@{\extracolsep{\fill}}ccc|cc|cc|cc@{}}
        \cline{1-9}
        \rule{0pt}{2.5ex}\textbf{Dataset} & \multirow{2}{1cm}{\textbf{Victim Model}} & \multirow{2}{1cm}{\textbf{Attack Method}}  
        & \multicolumn{2}{c|}{\textbf{Attack Effectiveness}} & \multicolumn{2}{c|}{\textbf{Attack Quality}}  & \multicolumn{2}{c}{\textbf{Our Improvement\%}}\\[0.05cm]
        \cline{4-9}
        & & & $\Delta_{drop}$ & \textbf{ASR\%} &  \rule{0pt}{2.5ex}\textbf{CAD} & \textbf{CodeBLEU} & \textbf{ASR\%} & $\Delta_{drop}$\\ [0.05cm]
        \cline{1-9}
        % 第一大栏
        \rule{0pt}{2.5ex}\multirow{9}{*}{Devign} & \multirow{4}{*}{CodeT5} & ALERT & 0.54 & 81.51 & {147.95} & {0.53} & - & -\\
            & & DIP & {0.41} & {53.05} & 54.66 & \cellcolor{pink!30}\textbf{0.86} & - & - \\
            & & HogVul  & \cellcolor{pink!30}\textbf{0.87} & \cellcolor{pink!30}\textbf{97.28} &  \cellcolor{pink!30}\textbf{316.45} & {0.84} & \textbf{+30.01} & \textbf{+0.395}\\[0.05cm]
        \cline{2-9}              
            \rule{0pt}{2.5ex} & \multirow{3}{*}{CodeBERT} & ALERT & 0.14 & 78.35 & {195.14} & {0.52} & - & -\\
            & & DIP & {0.21} & {59.75} & {41.02} & \cellcolor{pink!30}\textbf{0.88} & - & -\\
            & & HogVul  & \cellcolor{pink!30}\textbf{0.44} & \cellcolor{pink!30}\textbf{89.71} & \cellcolor{pink!30}\textbf{277.49} & {0.81} & \textbf{+20.66} & \textbf{+0.265}\\[0.05cm]
        \cline{2-9}
            \rule{0pt}{2.5ex} & \multirow{3}{*}{\centering GraphCodeBERT} & ALERT & 0.22 & 79.02 & {140.09} & {0.52} & - & -\\
            & & DIP & \cellcolor{pink!30}\textbf{0.29} & {49.45} & {38.93} & \cellcolor{pink!30}\textbf{0.86} & - & -\\
            & & HogVul  & {0.26} & \cellcolor{pink!30}\textbf{91.6} & \cellcolor{pink!30}\textbf{79.8} & {0.80} & \textbf{+40.42} & \textbf{+0.13}\\ [0.05cm]
        \cline{1-9}
    % 第二大栏
    \rule{0pt}{2.5ex} \multirow{9}{*}{DiverseVul} & \multirow{3}{*}{CodeT5} & ALERT & 0.28 & 59.98 & 36.66 & {0.29} & - & -\\
            &   & DIP & {0.5} & {74.53} & {18.25} & \cellcolor{pink!30}\textbf{0.92} & - & -\\
            &   & HogVul  & \cellcolor{pink!30}\textbf{0.61} & \cellcolor{pink!30}\textbf{96.50} & \cellcolor{pink!30}\textbf{210.9} & {0.84} & \textbf{+22.04} & \textbf{+0.29}\\ [0.05cm]
        \cline{2-9}
            \rule{0pt}{2.5ex} & \multirow{3}{*}{CodeBERT} & ALERT & 0.16 & 44.89 & {33.65} & {0.32} & - & -\\
            &   & DIP & {0.4} & {67.68} & 23.67 & 0.90 & - & -\\
            &   & HogVul  & \cellcolor{pink!30}\textbf{0.32} & \cellcolor{pink!30}\textbf{80.61} & \cellcolor{pink!30}\textbf{168.22} & \cellcolor{pink!30}\textbf{0.91} & \textbf{+15.95} & \textbf{+0.11}\\[0.05cm]
        \cline{2-9}
            \rule{0pt}{2.5ex} & \multirow{3}{*}{\centering GraphCodeBERT} & ALERT & 0.28 & 46.15 & {40.81} & {0.32} & - & -\\
            &   & DIP & {0.29} & {59.59} & 16.95 &  \cellcolor{pink!30}\textbf{0.9106} & - & -\\
            &   & HogVul  & \cellcolor{pink!30}\textbf{0.33} & \cellcolor{pink!30}\textbf{81.73} & \cellcolor{pink!30}\textbf{195.88} & {0.90}& \textbf{+35.02} & \textbf{+0.06}\\ [0.05cm]
    \cline{1-9}
    % 第三大栏
    \rule{0pt}{2.5ex} \multirow{9}{*}{BigVul} & \multirow{3}{*}{CodeT5} & ALERT & 0.13 & 48.14 & {58.06} & {0.31}& - & -\\
          &  & DIP & {0.39} & {72.32} & 39.08 & \cellcolor{pink!30}\textbf{0.9559} & - & -\\
          &  & HogVul  & \cellcolor{pink!30}\textbf{0.4} & \cellcolor{pink!30}\textbf{84.76} & \cellcolor{pink!30}\textbf{206.34} & {0.92}& \textbf{+23.05} & \textbf{+0.245}\\ [0.05cm]
        \cline{2-9} 
        \rule{0pt}{2.5ex} & \multirow{3}{*}{CodeBERT} & ALERT & 0.16   & {56.02} & {32.86} & {0.33}& - & -\\
        & & DIP & {0.38} & {67.65} & 22.86 & \cellcolor{pink!30}\textbf{0.98}& - & -\\
        & & HogVul  & \cellcolor{pink!30}\textbf{0.24} & \cellcolor{pink!30}\textbf{75.29} & \cellcolor{pink!30}\textbf{144.7} & {0.92}& \textbf{+24.36} & \textbf{+0.12}\\ [0.05cm]
        \cline{2-9}
        \rule{0pt}{2.5ex} & \multirow{3}{*}{GraphCodeBERT} & ALERT & 0.09 & {56.06} & {50.43} & {0.31}& - & -\\
        & & DIP & {0.13} & {56.00} & 18.55 & \cellcolor{pink!30}\textbf{0.99}& - & -\\
        & & HogVul  & \cellcolor{pink!30}\textbf{0.18} & \cellcolor{pink!30}\textbf{90.00} & \cellcolor{pink!30}\textbf{136.02} & {0.96}& \textbf{+24.36} & \textbf{+0.09
}\\ [0.05cm]
    \cline{1-9}
    \end{tabular*}
    \caption{The results present the attack effectiveness and attack quality of different attack methods (ALERT, DIP, HogVul) on various victim models for the datasets. Attack effectiveness is evaluated based on the model's performance drop ($\Delta_{drop}$) and attack success rate (ASR\%), while attack quality is measured by the code average diversity (CAD) and the CodeBLEU score.}
    \label{tab:code-code}
\end{table*}

\begin{figure*}
    \centering
    \includegraphics[width=1.0\textwidth]{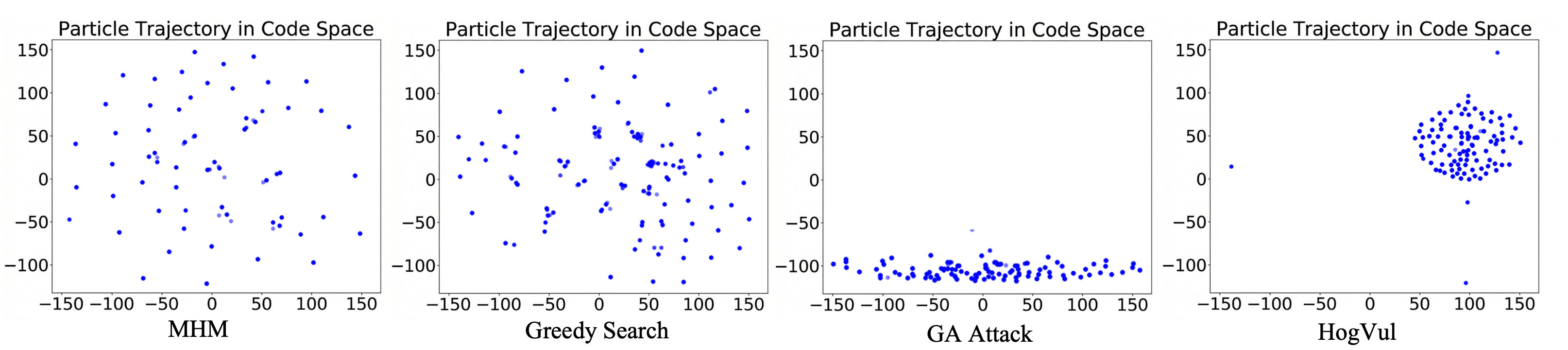}
    \caption{Visualization of particle trajectories in the code space during iterations of different optimization algorithms. }
    \label{fig:iter_trajectory}
\end{figure*}

To comprehensively evaluate our proposed method, we organize the analysis around three core experiments: \textit{overall attack performance}, \textit{design rationality}, and \textit{component effectiveness}.  
First, we compare HogVul with baseline methods across multiple metrics to assess its performance in terms of attack success, perturbation quality, and query efficiency.
Second, we investigate the design rationale of our optimization strategy by comparing PSO with alternative search algorithms.
Third, we conduct an ablation study to evaluate the contribution of each component within HogVul.
To further assess its robustness between the test sets and the training sets, we conduct a transferability experiment on the previously unseen dataset (Appendix E).
% ~\ref{sec:appendix_Generalization_Experiment}
Additionally, to assess HogVul’s applicability in real-world scenarios, we simulate attacks by measuring the false negative rate (FNR) (Appendix F).
% ~\ref{sec: real sene efficiency}             % ~\ref{sec:example}
Representative adversarial examples are provided in Appendix G to illustrate HogVul’s perturbation strategies.

% \subsection{RQ1: How does our method compare to baselines in terms of efficiency and quality?}
\subsection{Attack Effectiveness}
The experimental results summarized in Table~\ref{tab:code-code} validate the effectiveness of HogVul in vulnerability detection, as it consistently outperforms two widely-used baselines across four evaluation metrics. The key findings are summarized as follows:

\textbf{Performance of HogVul.}  HogVul consistently outperforms both baselines across all datasets. On average, it achieves a 26.37 higher ASR than ALERT and 25.72 higher than DIP, indicating stronger attack effectiveness. Furthermore, HogVul induces a larger model confidence drop, 0.18 compared to ALERT and 0.07 compared to DIP. This highlights the value of multi-level perturbations capability in solving the vulnerability detection task and also confirms the effectiveness of HogVul.

\textbf{Enhanced Perturbation Diversity.} In terms of perturbation diversity, HogVul generates a broader range of transformations than both baselines, as reflected by higher CAD scores. Although DIP reports higher CodeBLEU, this is primarily due to its reliance on inserting unreachable code, which minimally alters the original structure. DIP’s unchanged AST structure and data flow allow it to retain high $Match_{ast}$ and $Match_{df}$ scores, as the inserted dead code simply adds redundant nodes without disrupting the core structure. However, such transformations tend to be less diverse and may not reflect real-world attack tricks. By contrast, HogVul combines both semantic-preserving lexical perturbation and structure-altering code transformation, achieving a better balance between diversity and adversarial impact.

\subsection{Search Behavior Analysis}
To explain the rationality and advancement of our core PSO-driven method, we compare it with three representative search algorithms: Metropolis-Hastings MCMC (MHM), Greedy Search, and Genetic Algorithm (GA).
We use t-SNE~\cite{t-sne} to visualize the search trajectories of the optimal individuals at each iteration within the code space. As shown in Figure~\ref{fig:iter_trajectory}, PSO forms compact and convergent clusters over iterations, indicating stable convergence towards high-quality regions. In contrast, MHM exhibits scattered and irregular distributions, suggesting poor guidance in high-dimensional discrete spaces. The Greedy algorithm shows multiple discontinuous clusters, implying limited exploration and a tendency to converge prematurely. GA demonstrates directional but loosely distributed trajectories, covering broader regions without focusing on a singular optimum, which result in suboptimal convergence efficiency.

\begin{figure}[htbp]
    % \centering
    \includegraphics[width=\columnwidth,height=0.6\textheight, keepaspectratio]{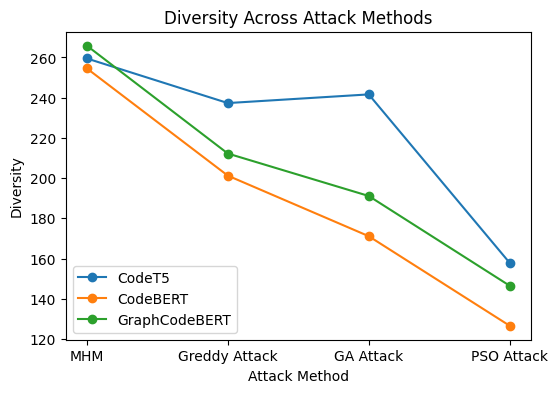}
    \caption{The diversity of the population in the last iteration.}
    \label{fig:diversity}
\end{figure}

To further validate the convergence behavior, we compute the population diversity at the final iteration using Eq.\ref{eq:diversity-metric}. As shown in Figure~\ref{fig:diversity}, PSO achieves the lowest diversity among all methods, reflecting its ability to concentrate the search around promising candidates.

\subsection{Query Efficiency Evaluation}
To quantify query efficiency, we propose a novel metric: \textit{ASR per Query (APQ)}, defined as the attack success rate normalized by the number of model queries. This metric provides a comprehensive measure of how effectively each query contributes to successful
%~\ref{sec:appendix_query_efficiency}
adversarial generation. As detailed in Appendix C, HogVul achieves the highest APQ in six out of nine scenarios, and remains competitive in the others. Although HogVul operates over a larger perturbation space, its higher APQ indicates that additional queries are strategically utilized to achieve more successful attacks, rather than wasted on redundant exploration.

\subsection{Practical Evaluation}
To evaluate HogVul under practical conditions, we conduct an experiment on unseen datasets from the D2A benchmark, which includes real-world vulnerabilities and diverse code styles that differ from the training data (Appendix E). This distribution shift simulates practical 
% ~\ref{sec:appendix_Generalization_Experiment}
scenarios where models encounter unfamiliar patterns, yet HogVul maintains high ASR without parameter tuning.                             %~\ref{sec: real sene efficiency}
Additionally, we simulate real-world deployment settings by measuring the False Negative Rate (FNR) when HogVul is applied to truly vulnerable code samples (Appendix F). Results show that HogVul consistently induces misclassification, particularly in four critical cases, underscoring its threat to vulnerability detectors.

\subsection{Ablation Study}
\begin{figure}
    \centering
    \includegraphics[width=\columnwidth]{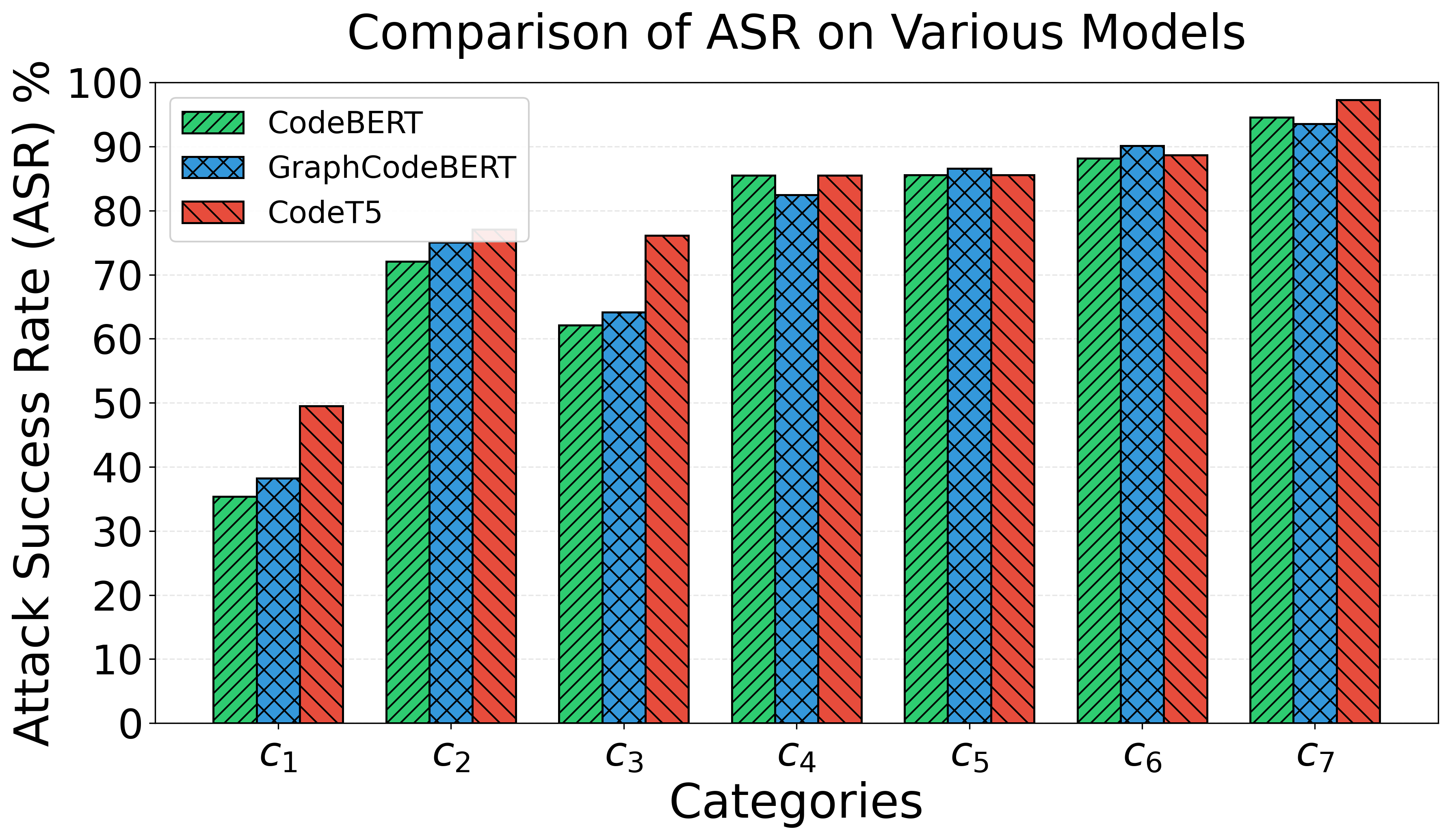}
    \caption{Comparison of ASR across component settings.}
    \label{fig:rq31}
\end{figure}

\begin{figure}
    \centering
    \includegraphics[width=0.5\textwidth]{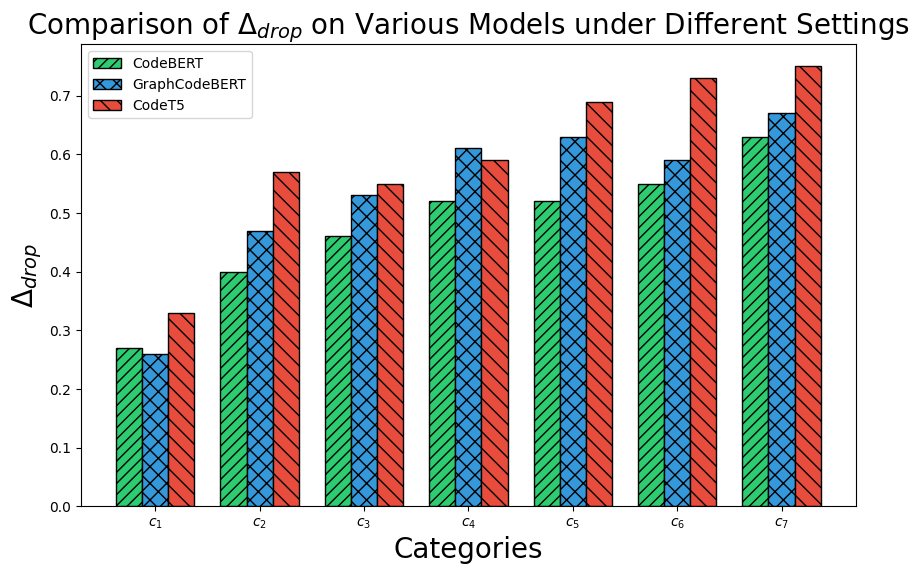}
    \caption{Comparison of  $\Delta drop$ across component settings.}
    \label{fig:rq32}
\end{figure}

\textbf{Effectiveness and Interaction of Key Components in HogVul.} 
We conduct an ablation study comprising seven configurations ($C_1$–$C_7$), detailed in Table~\ref{tab:rq3_config}. These configurations progressively remove or combine key modules, with $C_7$ serving as the full baseline model. The results in Figure~\ref{fig:rq31} and Figure~\ref{fig:rq32} show that removing lexical-level perturbation ($C_2$) significantly reduces ASR, underscoring that PSO plays a critical role in guiding efficient and targeted exploration. Similarly, removing structural transformations ($C_3$) lowers $\Delta_{drop}$ and ASR, emphasizing the contribution of syntax-level perturbation to robust adversarial generation.
Combining any two modules ($C_4$–$C_6$) notably improves performance, indicating the components reinforce each other during the adversarial generation process. The baseline configuration ($C_7$) achieves the best ASR (97.28\% on CodeT5), illustrating the effectiveness of all components working together. These findings confirm that each component contributes distinct and complementary capabilities.

\begin{table}[h!]
\centering
\resizebox{\columnwidth}{!}{ 
    \begin{tabular}{c c}
    \toprule 
    \textbf{Configuration} & \textbf{Description} \\
    \midrule
    $C_1$ & Only Semantic-Aware Initiation\\
    $C_2$ & Only Lexical Perturbation (Random Init) \\
    $C_3$ & Only Structure-Aware Transformation \\
    $C_4$ & $C_1$ + $C_2$ \\
    $C_5$ & $C_1$ +$C_3$ \\
    $C_6$ & $C_2$+ $C_3$ (Random Init + Cooperative Optimization) \\
    $C_7$ &$C_1$ +$C_2$ + $C_3$  \\
    % \hline
    \bottomrule
    \end{tabular}
}
\caption{The configuration of ablation study.}
\label{tab:rq3_config}
\end{table}

\section{Conclusion}
\label{sec:conclusion}

In this paper, we present HogVul, a black-box adversarial attack framework for LM-based vulnerability detection that integrates both lexical and syntax perturbations under a unified dual-channel optimization. HogVul achieves high attack success rates while maintaining functional correctness.
Extensive experiments across multiple models and datasets validate its effectiveness and robustness.
Our findings underscore the potential of hybrid optimization strategies in exposing vulnerabilities and emphasize the necessity of robust defenses against adversarial attacks in vulnerability detection. 
Furthermore, the modular nature of HogVul allows for potential integration with other attack strategies, paving the way for future research on hybrid adversarial attack frameworks. Future work will include seeking more reasonable search paradigms and theoretical support.

\clearpage
\section*{Acknowledgements}
\label{sec:Acknowledge}

This work was partly supported by the National Key Research and Development Program of China under No. 2024YFB3908400, NSFC under No. 62402418, the Zhejiang Province's 2025 "Leading Goose + X” Science and Technology Planunder grant No.2025C02034, the Key R\&D Program of Ningbo under No. 2024Z115.

We thank the support from the Zhejiang University College of Computer Science and Technology, the Zhejiang University NGICS Platform and the Qiushi Flying Eagle Project from Zhejiang University.

% \bibliography{aaai2026}

\clearpage
\section*{Appendix}
\label{sec:appendix}

\subsection*{A The Details of PSO-Driven Lexical Perturbation}
\label{sec:design of PSO}

\begin{algorithm}[t]
\caption{Structure-Aware Code Transformation}
\label{algo:structure-aware}
\begin{algorithmic}[1] % [1] 表示每行都编号
    \REQUIRE Source code: $c$, Model: $M$
    \ENSURE Adversarial example: $c'$
    \STATE $S \gets \texttt{extract\_structural}(c)$
    \STATE $F \gets \texttt{compute\_frequencies}(S)$
    \FORALL{structure type $i$ in $S$}
        \IF{$i$ is control flow structure}
            \STATE $\phi(i) = \lambda \cdot F[i]$
        \ELSE
            \STATE $\phi(i) = F[i]$
        \ENDIF
    \ENDFOR
    \STATE $P = \texttt{normalize\_distribution}(\phi)$
    \STATE $c_{\text{masked}} = \texttt{get\_masked\_tokens}(c)$
    \STATE $I = \texttt{compute\_importance}(M, c_{\text{masked}})$
    \STATE $\text{selected} = \texttt{sample\_structures}(P)$
    \STATE $\text{positions} = \texttt{get\_positions}(I)$
    \STATE $c' = \texttt{apply\_trans}(c, \text{selected}, \text{positions})$
    \RETURN $c'$
\end{algorithmic}
\end{algorithm}

This section provides a comprehensive explanation of the PSO-driven lexical perturbation strategy in HogVul, including its directional guidance mechanisms, adaptive parameter scheduling, probabilistic update strategy, and diversity maintenance.

\noindent \textbf{Directional Guidance in Lexical Substitution Space.}
Each particle represents a candidate adversarial code, encoded as a vector $X_t$ of variable renamings, where $t$ is an iterative round. Each dimension corresponds to an identifier in the source code and its value specifies the current substitute.

The evolution of particles is governed by three directional components:
\textit{Inertia} retains a fraction of the particle’s previous directional movement, promoting stability in the search trajectory.
\textit{Individual best} directs each particle toward its own historically most effective renaming configuration, reinforcing previously successful perturbations unique to that particle. This component encourages particles to explore and refine their personalized attack strategies, maintaining diversity within the swarm.
\textit{Global best} guides all particles toward the globally most effective configuration observed across the entire swarm, promoting convergence toward the optimal perturbation that has induced the largest model confidence drop. This collective guidance ensures that the swarm exploits proven attack patterns, sharing the optimal perturbation information within the swarm.

These components are combined to compute a velocity vector, which suggests new renaming choices in each dimension. For example, if a particle previously substituted \texttt{copied} $\rightarrow$ \texttt{read\_bytes}, achieving a moderate confidence drop. Meanwhile, another particle’s best result used \texttt{copied} $\rightarrow$ \texttt{total\_read}, causing a greater model confidence drop. Under the influence of the global best, the particle’s velocity vector will be adjusted to increase the probability of adopting \texttt{total\_read} as the new substitution for \texttt{copied}. As particles update their positions accordingly, they generate new code variants that gradually converge toward adversarially effective perturbations.

\noindent \textbf{Adaptive Parameter Scheduling.}
To balance global exploration and local exploitation during the optimization process, we adopt a time-dependent scheduling strategy for PSO's hyperparameters. The goal is to allow particles to explore the perturbation space more freely in the early stages, and gradually focus on exploiting promising regions as the optimization proceeds.

Specifically, the \textit{inertia weight} $\omega(t)$ controls how much of the particle’s previous velocity is retained in the current update. A higher inertia weight encourages exploration, while a lower one favors convergence. We adopt a linear decay strategy:
\begin{equation}
\omega(t) = (\Omega_1 - \Omega_2) \cdot \frac{T - t}{T} + \Omega_2
\end{equation}
where $\Omega_1 = 1.5$ and $\Omega_2 = 0.6$ denote the initial and final inertia weights, and $T$ is the total number of iterations. This schedule gradually reduces the momentum, encouraging convergence in later stages.

The \textit{cognitive} ($C_1$) and \textit{social} ($C_2$) acceleration coefficients determine how strongly a particle is pulled toward its personal best and the global best position, respectively. Their scheduling ensures a smooth transition from self-exploration to swarm consensus:
\begin{equation}
\begin{gathered} 
C_{1}(t)=C_1^{\mathrm{ori}}-\frac{t}{T}\cdot(C_{1}^{\mathrm{ori}}-C_2^{\mathrm{ori}}) \\
C_2(t)=C_2^{\mathrm{ori}}+\frac{t}{T}\cdot(C_1^{\mathrm{ori}}-C_2^{\mathrm{ori}})
\end{gathered}
\label{eq:update}
\end{equation}
Here, $C_1^{\mathrm{ori}}$ and $C_2^{\mathrm{ori}}$ are initial values, We set them to 1.3 and 0.6 respectively to encourage individual exploration at the start and coordinated convergence later.

With these schedules in place, the velocity update rule for each particle becomes:
\begin{equation}
\begin{aligned}
V_{t+1} &= \omega(t)V + C_1(t)(G_{\text{local}} - X_t) \\
&\quad + C_2(t)(G_{\text{global}} - X_t)
\end{aligned}
\label{eq:velocity}
\end{equation}
This equation models the directional influence of momentum, personal memory and social learning on each particle’s movement across the lexical perturbation space.

\noindent \textbf{Probabilistic Dimension Update.}
Since lexical substitution is discrete, directly applying real-valued velocity updates is infeasible. To address this, we adopt a probabilistic update scheme that translates velocity magnitude into a substitution likelihood. Specifically, for each variable dimension $d$:
\begin{equation}
P(\text{update}) = \sigma(V_{id}) = \frac{1}{1 + e^{-V_{id}}}
\end{equation}
This sigmoid function interprets the velocity as a soft indicator of whether a variable should be updated. Dimensions with higher directional momentum are more likely to be perturbed, enabling adaptive and data-driven exploration.

\noindent \textbf{Diversity Maintenance and Stagnation Handling.}
To avoid premature convergence and enhance diversity, we introduce two mechanisms: 
(1) Random mutation with a small probability $p_{\text{mutate}}$ is applied to particles at each step, introducing stochasticity and encouraging diversity.
(2) We maintain a stagnation counter for the global best solution. If no improvement is observed over $T$ consecutive iterations, we trigger our \textit{dual-channel optimization mechanism}, which  will be explained in Section~\ref{sec:section 3.5}.
\noindent In summary, these adaptive schedules and discrete adaptations ensure that PSO can be effectively deployed in the adversarial code perturbation setting, guiding the swarm toward high-impact, semantically consistent lexical attacks.

\begin{figure*}[t]
    \centering
    \begin{subfigure}[b]{0.3\linewidth}
        \centering
        \includegraphics[width=\linewidth]{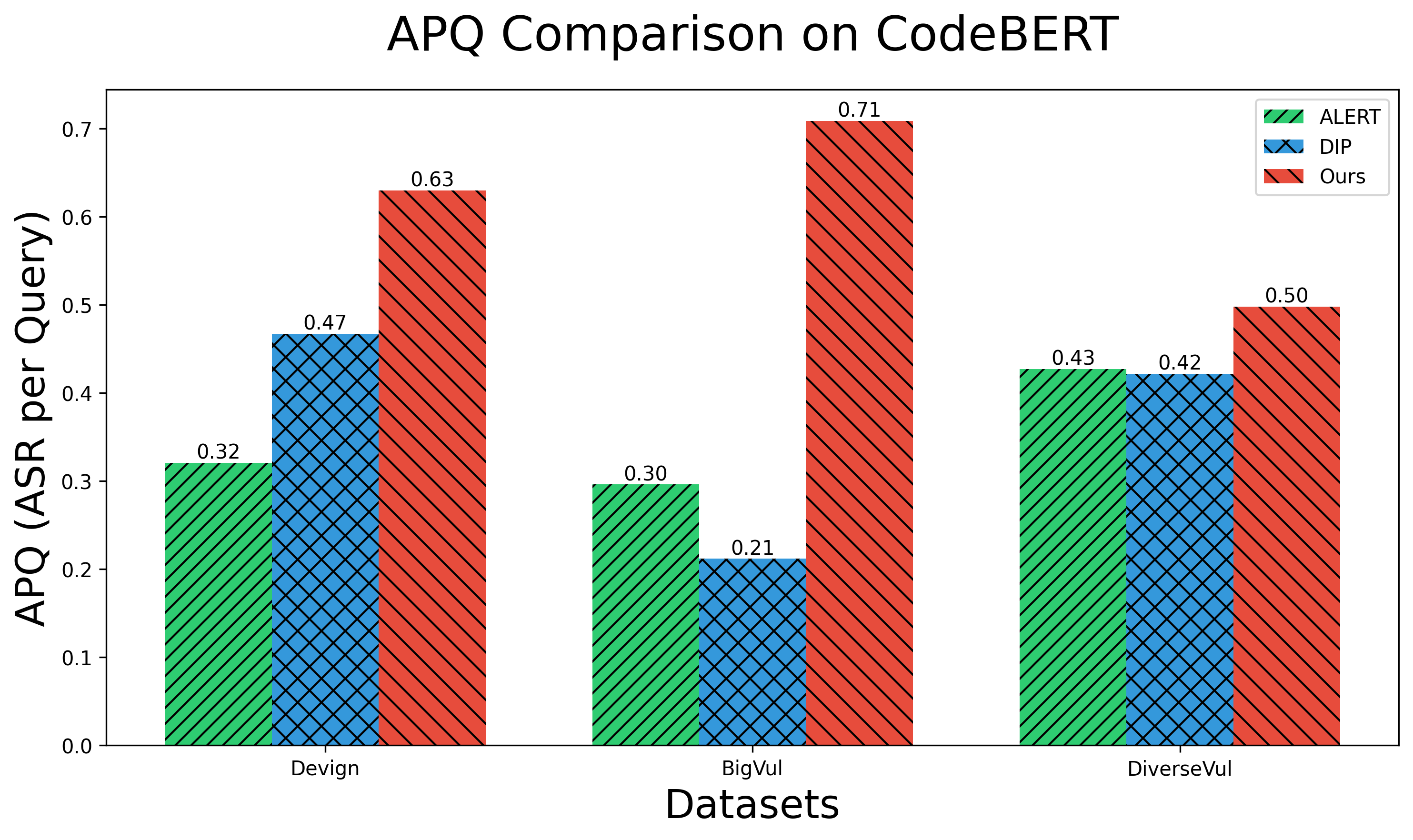}
        \caption{CodeBERT}
    \end{subfigure}
    \hfill
    \begin{subfigure}[b]{0.3\linewidth}
        \centering
        \includegraphics[width=\linewidth]{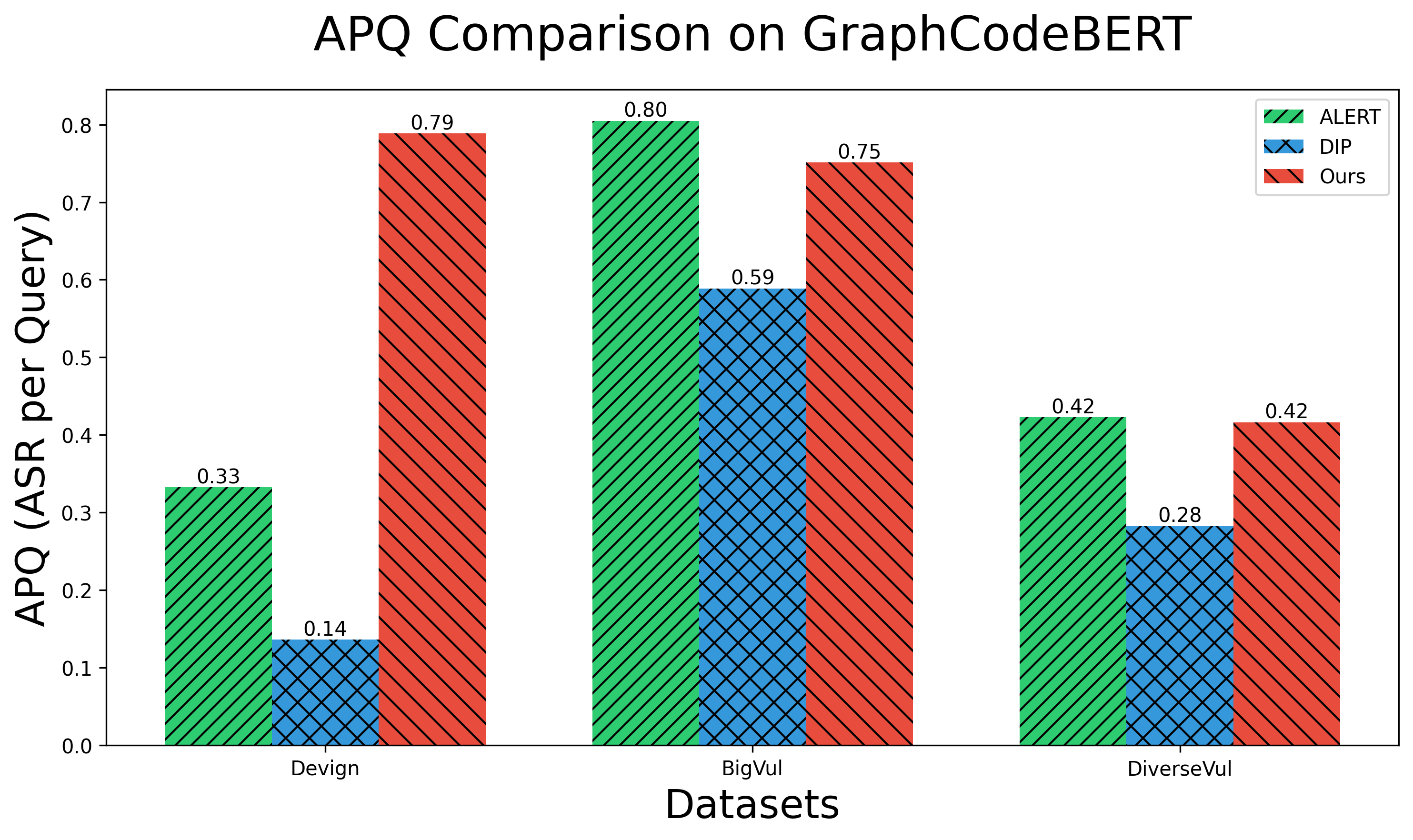}
        \caption{GraphCodeBERT}
    \end{subfigure}
    \hfill
    \begin{subfigure}[b]{0.3\linewidth}
        \centering
        \includegraphics[width=\linewidth]{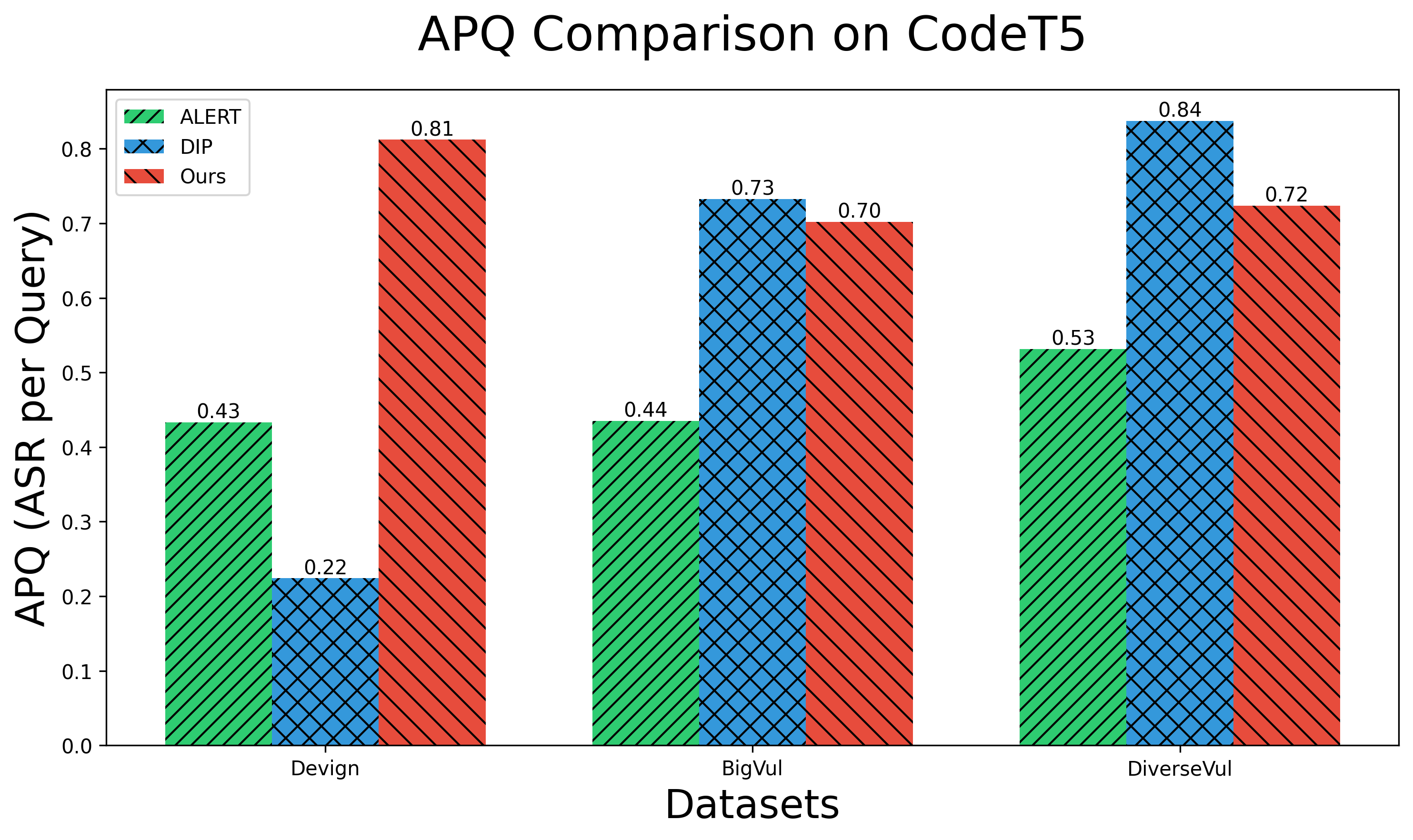}
        \caption{CodeT5}
    \end{subfigure}

    \caption{Comparison of ASR per query (APQ) on different models and different datasets.}
    \label{fig:asr_per_query}
\end{figure*}

\subsection*{B Experiments Setup}

\subsubsection{B.1 Datasets}
\label{sec:appendix_datasets}
In this section we provide detailed descriptions of the benchmark datasets used in our experiments.

\textbf{Devign}~\cite{devign}: Consists of 48,000 functions extracted from two popular open-source C projects: FFmpeg and Qemu. This dataset primarily focuses on learning representations for vulnerability detection. 

\textbf{DiverseVul}~\cite{diversevul}: A large-scale and diverse C/C++ vulnerability dataset, containing 18,945 vulnerable and 330,492 non-vulnerable functions from 797 projects spanning 150 CWE categories. We sample 20,000 instances from the first ten projects, ensuring a balanced 1:1 ratio between vulnerable and non-vulnerable functions.

\textbf{BigVul}~\cite{bigvul}: An extensive C/C++ vulnerability dataset comprising 3,754 vulnerabilities collected from 348 open-source projects, spanning 91 vulnerability types. BigVul focuses on diverse code bases and includes real-world vulnerabilities reported in publicly available software repositories. This dataset emphasizes structural and logical vulnerabilities, providing a robust testbed for evaluating the robustness of adversarial attack strategies.

\textbf{D2A}~\cite{D2A}: Consists of over 1.3 million labeled examples generated from multiple open-source C/C++ projects, including OpenSSL, FFmpeg, httpd, NGINX, libtiff and libav. We selected OpenSSL and NGINX from the D2A dataset for generalization testing, as they are absent from the previous datasets and represent distinct vulnerability distributions and code styles.

\subsubsection{B.2 Baseline Methods}
\label{sec:appendix_baseline}

In our study, we chose DIP \cite{dip} and ALERT \cite{alert} as baseline methods to evaluate our adversarial attack approach. In the area of vulnerability detection, these two approaches represent two different but complementary strategies for attacking LM-based methods. 

ALERT focuses on lexical-level perturbation, leveraging pre-trained models to generate context-aware substitutions to preserve the naturalness of adversarial examples, ensuring they are functionally correct and human-readable. By integrating greedy search and genetic algorithms, ALERT effectively navigates discrete code space while maintaining readability and context consistency. 

In contrast, DIP operates at the syntax-level, introducing semantically neutral dead code fragments to exploit the structural sensitivity of language model-based vulnerability detectors. DIP ensures the adversarial examples remain compilable, making them highly applicable in real-world scenarios, while achieving a near-perfect CodeBLEU score, indicating minimal semantic drift. Both methods have been extensively evaluated on CodeBERT, GraphCodeBERT, and CodeT5, which are also the LM-based detectors used in our study. 

Both ALERT and DIP are widely recognized and publicly available, providing complementary perturbation strategies that serve as robust baselines for evaluating HogVul, allowing us to rigorously evaluate our approach against a variety of adversarial attack tactics. By including these two methods, we provide a balanced and rigorous benchmark to validate the multi-level perturbation capability of HogVul.

\subsubsection{B.3 Evaluation Metrics}
\label{sec:mertics}

To comprehensively evaluate the performance of HogVul in generating adversarial code while preserving functional correctness, we adopt five core metrics aligned with our objectives: attack effectiveness and code quality retention.

\textbf{Attack Success Rate (ASR)} quantifies the proportion of adversarial examples that successfully induce misclassification by the target model:
\begin{equation}
\text{ASR} = \frac{\text{Successful Attacks}}{\text{Total Attempts}} \times 100\%
\end{equation}
A higher ASR indicates stronger adversarial impact and highlights the vulnerability of the target model to such attacks.

\textbf{ASR per Query (APQ)} is a novel metric we propose to assess attack efficiency. It normalizes the success rate by the average number of model queries required per successful attack:
\begin{equation}
\text{APQ} = \frac{\text{ASR\%}}{\text{Avg Queries per Attack}}
\label{eq:apq}
\end{equation}
This metric reflects the query cost of achieving successful adversarial perturbations, enabling fair comparisons under differing query budgets and search complexities.

\textbf{CodeBLEU}~\cite{CodeBLEU} evaluates the similarity between adversarial and original code by combining syntactic and semantic metrics:
\begin{equation}
\begin{aligned}
\text{CodeBLEU} = &\alpha \cdot \text{BLEU} + \beta \cdot \text{BLEU}_{\text{weight}} \\
&+ \gamma \cdot \text{Match}_{\text{ast}} + \delta \cdot \text{Match}_{\text{df}}
\end{aligned}
\end{equation}
\noindent where $BLEU$ measures the n-gram overlap between the generated code and the original code; $BLEU_{weight}$ emphasizes semantic keywords; $Match_{ast}$ evaluates syntactic similarity via abstract syntax trees and $Match_{df}$ captures dataflow preservation. We follow prior work and set all weights $\alpha, \beta, \gamma, \delta$ to 0.25.

\textbf{Average Confidence Drop ($\Delta_{\text{drop}}$)} captures the average decrease in the model's confidence for the correct label after applying adversarial perturbations:
\begin{equation}
\Delta_{\text{drop}} = \frac{1}{N}\sum_{i=1}^N (C_{\text{orig}}^{(i)} - C_{\text{adv}}^{(i)})
\end{equation}
\noindent where $C_{\text{orig}}$ and $C_{\text{adv}}$ denote the model's predicted probability for the correct class on original and adversarial examples, respectively. A higher drop implies stronger confusion introduced by the perturbation.

\textbf{Code Average Diversity (CAD)} is another novel metric we propose to quantify population-level perturbation diversity. It computes the average pairwise normalized Levenshtein distance~\citep{haldar2011levenshteindistancetechniquedictionary} among substitution vectors:
\begin{equation}
\text{CAD} = \frac{2}{n(n-1)}\sum_{i=1}^{n-1} \sum_{j=i+1}^n \text{Levenshtein}(p_i, p_j)
\label{eq:diversity-metric}
\end{equation}
\noindent A higher CAD indicates broader lexical coverage and greater diversity in adversarial strategies, which is desirable for exploring a wider attack surface and avoiding overfitting to specific perturbation patterns.

\subsubsection{B.4 Hyperparameters}
\label{sec:Hyperparameters}

\begin{table}[t]
\centering
% \caption{The hyperparameter settings of our method.}
% \label{tab:hyperparameters}
\begin{tabular}{c c}
\hline
\textbf{Hyperparameter} & \textbf{Value} \\
\hline
Population size ($pop\_size$) & 20 \\
Max iterations ($max\_iters$) & 20 \\
Initial inertia weight ($\omega_1$) & 1.5 \\
Final inertia weight ($\omega_2$) & 0.6 \\
Personal learning factor ($C_1^{ori}$) & 1.3 \\
Social learning factor ($C_2^{ori}$) & 0.6 \\
Stagnation threshold ($\theta$) & 2 \\
Mutation probability ($p_{mutate}$) & 0.1 \\
\hline
\end{tabular}
\caption{The hyperparameter settings of our method.}
\label{tab:hyperparameters}
\end{table}

The hyperparameters of our method are set as Table~\ref{tab:hyperparameters}. We configure a population size of 20 and set the maximum number of iterations to 20. The inertia weight is linearly decayed from an initial value of 1.5 to a final value of 0.6. The personal and social learning factors are set to 1.3 and 0.6, respectively. Additionally, a stagnation threshold of 2 is introduced to prevent premature convergence, while a mutation probability of 0.1 enhances diversity in the search space, increasing the likelihood of discovering effective adversarial perturbations.

\subsection*{C Query Efficiency Analysis}
\label{sec:appendix_query_efficiency}  
To further evaluate the efficiency of our attack, we introduce the Query Normalized Attack Success Rate (ASR per Query, APQ), which quantifies the attack success rate relative to the number of query attempts. Comparing the total number of queries alone may not accurately reflect efficiency without considering the impact of the search space. To address this limitation, we define APQ in Eq~\ref{eq:apq}.

To ensure a fair comparison, we conducted experiments on two baseline methods and our proposed method using the same three datasets and three target models. The results are shown in Figure~\ref{fig:asr_per_query}, our approach achieves the highest APQ in six scenarios, outperforming all baseline methods. In the remaining three scenarios, it closely approaches the optimal APQ value. Notably, despite operating within a broader search space, our method consistently attains a significantly higher ASR, demonstrating that additional queries contribute meaningfully to attack success rather than being redundant.  

These results highlight the effectiveness of our approach in generating adversarial examples, ensuring that each query is optimally utilized to enhance attack performance. By maximizing attack efficiency while maintaining high ASR, suggesting that our hybrid optimization strategy not only improves overall attack performance, but also offers better cost-effectiveness in black-box settings. The consistent improvements in APQ further underscore the practical value of incorporating query-aware optimization into adversarial code generation.

\subsection*{D Control Flow Weight Ablation Experiment}
\label{sec:ablation experiment}
It has been shown that the deep attention layers of code language models exhibit significant sensitivity to control flow structures~\citep{naturalcc,syntaxguide}. Therefore, increasing the disturbance weight of control flow structures can more effectively disrupt the model’s recognition of key structural patterns.  

\begin{figure}
    \centering
    \includegraphics[width=0.5\textwidth]{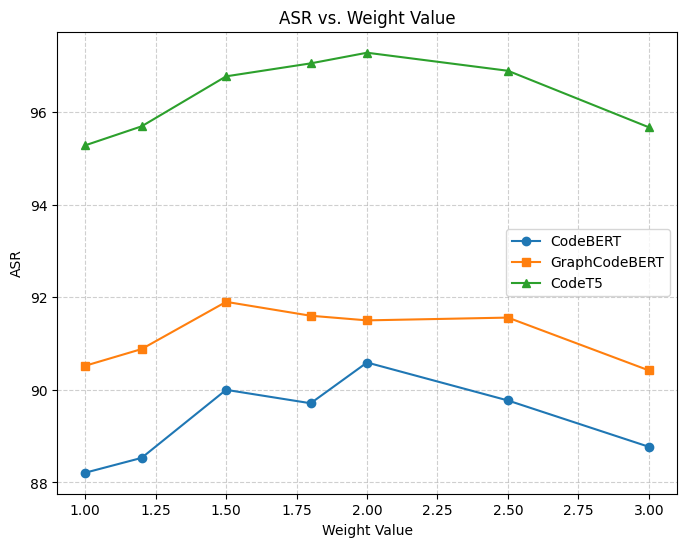}
    \caption{Attack Success Rate (ASR\%) under different weights}
    \label{fig:weight_compare}
\end{figure}

To evaluate the impact of control flow weight in Eq~\ref{eq:probability-distribution} on attack effectiveness, we conduct an ablation study with the following settings: The control-flow weight is varied across \{1, 1.2, 1.5, 1.8, 2, 2.5, 3\}, while the weights for other structures are fixed at 1.0. The results, presented in Figure~\ref{fig:weight_compare}, reveal a clear trend: As the control flow weight increases from 1.0 to 2.0, the attack success rate (ASR) improves consistently, reaching a peak at 1.8. However, beyond this threshold (e.g., 2.5 and above), ASR declines, indicating a diminishing effect.  

These findings suggest that moderate perturbation of control flow structures enhances attack effectiveness, whereas excessive perturbation may reduce its impact. We speculate that this is due to the high sensitivity of code language models to control flow structures: perturbing within a certain range effectively disrupts key pattern recognition, improving ASR. However, excessive disturbance may exceed the model’s generalization capacity, causing it to disregard portions of the perturbed structure, ultimately undermining the attack’s effectiveness.

\begin{table*}[t]
\scriptsize
\newcommand{\tabincell}[2]{\begin{tabular}{@{}#1@{}}#2\end{tabular}}  
  \centering
  \begin{tabular} {p{3cm}p{5cm}p{4.2cm}p{4.2cm}}
\toprule

\textbf{Selected Control Flow} & \tabincell{c}{\textbf{Description}} &  \tabincell{c}{\textbf{Before}} & \tabincell{c}{\textbf{After}} \\
\midrule

Op1-For2While & The for-loop is transformed into a while-loop. & \tabincell{l}{    for(i=0;i<10;i++)\{\\ \quad  BodyA\\\}  } &  \tabincell{l}{ i=0; while(i<10)\{ \\ \quad BodyA \\ \quad i++; \} }
 \\
\hline
Op2-While2For &The while-loop is transformed into  a for-loop. & \tabincell{l}{while(i<10)\{ \\ \quad BodyA \} } &\tabincell{l}{    for(;i<10;)\{\\ \quad BodyA\}  } \\
\hline

 Op3-ChIfElse2Else  & Transformation of if elseif to if else. & \tabincell{l}{ if(grad<60)  BodyA\\ else if(grad<80)   BodyB \\ else  BodyC } & \tabincell{l}{if(grad<60) BodyA \\ else\{    if(grad<80)   BodyB \\  \quad else  BodyC \}}\\
\hline
Op4-ChElse2ElseIf & Transformation of if else to if elseif. & \tabincell{l}{if(grad<60) BodyA \\ else\{ if(grad<80)   BodyB \\  \quad else  BodyC \}} &\tabincell{l}{ if(grad<60)  BodyA\\ else if(grad<80)   BodyB \\ else  BodyC }  \\
\hline
Op5-ChSwitch &  Transformation of the  Switch statement to the  if elseif statement. & \tabincell{l}{switch(a)\{  case 60: BodyA \\ \quad case 70: BodyB\\ \quad default: BodyC \}} & \tabincell{l}{if(a==60) BodyA\\ else if(a==70) BodyB\\ else BodyC }\\
\hline
Op6-ChDo & The do-loop is transformed into a while-loop. & \tabincell{l}{do\{\\ \quad BodyA \\\}while(i<10);} & \tabincell{l}{BodyA\\while(i<10)\{ \\\quad BodyA\}
}\\
\bottomrule
\end{tabular}
% \vspace{-2mm}
\caption{The descriptions and examples of selected control flow}
\label{tab:control_flow_example}
\end{table*}

\subsection*{E Practical Evaluation Under Distribution Shift of dataset}
\label{sec:appendix_Generalization_Experiment}

To assess HogVul’s effectiveness in realistic deployment scenarios, we conduct experiments on a previously unseen dataset that exhibit distribution shifts compared to the original training data. This evaluation simulates practical conditions where the attack setting differs from the training context, allowing us to examine HogVul’s robustness against diverse code styles and vulnerability patterns.

We utilize the D2A dataset~\cite{D2A}, specifically focusing on the OpenSSL and NGINX projects. These projects are excluded from the original evaluation datasets but represent realistic, large-scale software systems. This setup provides an opportunity to assess HogVul’s performance when encountering new codebases and vulnerability types. To ensure fair comparison, all experimental parameters are maintained consistent with the original setup, isolating the impact of data distribution differences.
\begin{table}[h]
    \centering
    % \caption{Performance of different methods on D2A Datasets}
    % \label{tab:Generalization Experiment}
    \resizebox{\columnwidth}{!}{
    \begin{tabular}[]{l l c c}
        \toprule
        Victim Model    & Attack Method & $\Delta_{drop}$ & ASR\% \\ 
        \midrule
        CodeT5          & ALERT         & 0.36           & 68.40 \\ 
                        & DIP           & \cellcolor{pink!30}\textbf{0.51}            & 83.51 \\ 
                        & HogVul        & {0.47}   & \cellcolor{pink!30}\textbf{95.79} \\ 
        \midrule
        CodeBERT        & ALERT         & 0.06           & 48.40 \\ 
                        & DIP           & 0.27            & 71.70 \\ 
                        & HogVul        & \cellcolor{pink!30}\textbf{0.38}   & \cellcolor{pink!30}\textbf{89.94} \\ 
        \midrule
        GraphCodeBERT   & ALERT         & 0.11           & 59.44 \\ 
                        & DIP           & 0.14           & 35.56 \\ 
                        & HogVul        & \cellcolor{pink!30}\textbf{0.42}   & \cellcolor{pink!30}\textbf{91.11} \\ 
        \bottomrule
    \end{tabular}
    }
    \caption{Performance of different methods on D2A Datasets}
    \label{tab:Generalization Experiment}
\end{table}
Table~\ref{tab:Generalization Experiment} presents the results. HogVul achieves the highest \textit{Attack Success Rate} (ASR\%) across all victim models: 95.79\% on CodeT5, 89.94\% on CodeBERT, and 91.11\% on GraphCodeBERT, notably outperform both ALERT and DIP.

Beyond ASR, we also evaluate the \textit{Average Confidence Drop} ($\Delta_{\text{drop}}$), which reflects the model’s reduction in certainty for the correct label after adversarial perturbation. HogVul induces the largest confidence drop on CodeBERT (0.38) and GraphCodeBERT (0.42), while its performance on CodeT5 (0.47) is close to that of DIP (0.51).

Notably, DIP and ALERT both suffer significant performance degradation in this setting. We attribute this to their stronger reliance on dataset-specific priors or learned substitution heuristics, which do not transfer well across unseen projects. In contrast, HogVul’s adaptive search strategy, combining lexical and structural perturbations, demonstrates strong transferability across varying data distributions, underscoring its robustness in practical scenarios.

Overall, the findings confirm that HogVul effectively adapts to distribution shifts without additional tuning, achieving high attack success on diverse and previously unencountered codebases. This reinforces its potential impact as a practical adversarial tool in real-world vulnerability detection systems.

\subsection*{F Real Scene Efficiency}
\label{sec: real sene efficiency}
To better evaluate the practical impact of adversarial attacks, we conduct a real-world simulation that focuses on the model's ability to detect actual vulnerable code. Existing evaluations typically mix secure and vulnerable samples and report overall Attack Success Rate (ASR), which may not fully reflect the consequences in realistic usage. In real deployment scenarios, false negatives vulnerabilities being missed will pose a greater security risk than false positives~\cite{falsenegatives}.

To address this, we design an experiment that exclusively targets real vulnerable samples and measures the \textit{False Negative Rate (FNR\%)} as the evaluation metric. Specifically, we apply HogVul to samples confirmed to contain vulnerabilities, sourced from three benchmark datasets used in our prior experiments. The goal is to assess whether HogVul can cause these genuinely vulnerable instances to be misclassified as safe, thereby revealing its ability to undermine model reliability in realistic settings.

\begin{figure}
    % \centering
    \includegraphics[width=0.5\textwidth]{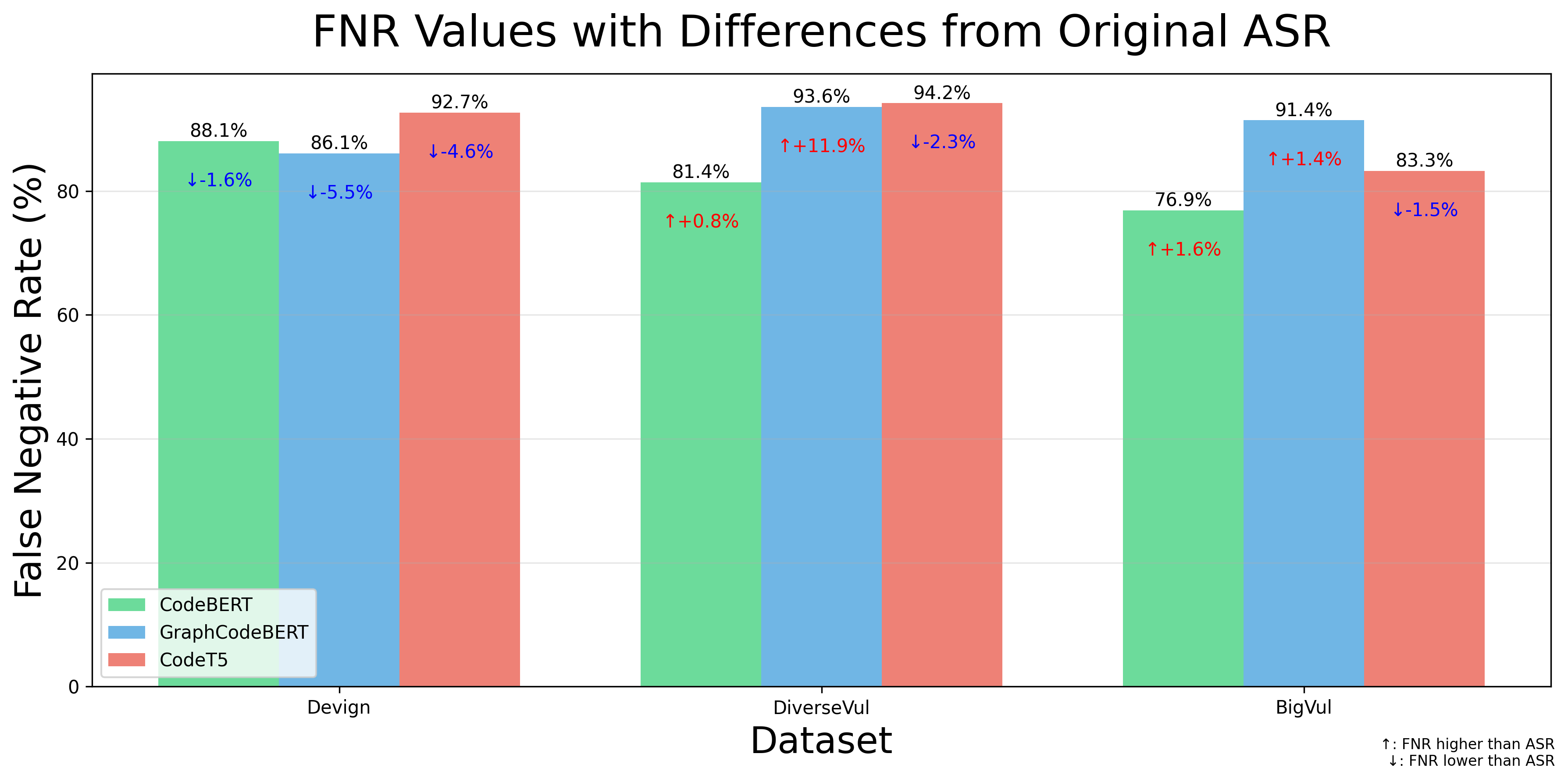}
    \caption{The comparison of the false negative rate (FNR\%) of the HogVul model in real scenarios with Table~\ref{tab:code-code}.}
    \label{fig:fnr_with_differences}
\end{figure}

The results are shown in Figure~\ref{fig:fnr_with_differences}, indicate that HogVul consistently increases the FNR across all test scenarios. In particular, it achieves notable increases in four representative subsets, demonstrating stable and impactful attack performance. These findings suggest that even subtle perturbations introduced by HogVul can compromise the detection of real-world vulnerabilities, highlighting the potential risks such attacks pose to the dependability of deployed vulnerability detection systems.

\begin{table*}[t]
    \centering
    \begin{tabular}{@{} l ccccc @{}}
        \toprule
        \textbf{Victim Model} 
            & \textbf{ITGen}$^*$ 
            & \textbf{SLODA} 
            & \textbf{GraphCodeAttack}$^*$ 
            & \textbf{CODA} 
            & \textbf{HogVul} \\
        \midrule
        CodeBERT       & 76.07 & \textbf{91.52} & 78.62 & 83.27 & 87.79 \\
        GraphCodeBERT  & 81.63 & 89.11          & 79.90 & 86.54 & \textbf{91.60} \\
        CodeT5         & 96.25 & --             & --    & 95.02 & \textbf{97.28} \\
        \midrule
        \textbf{Average} & 84.65 & 90.31 & 79.26 & 88.27 & \textbf{92.22} \\
        \bottomrule
    \end{tabular}
    \caption{Attack Success Rate(ASR\%) of HogVul and contemporaneous baselines on the Devign dataset. $*$ indicates results reported in the original paper.}
    \label{tab:broader_baselines}
\end{table*}

\begin{table*}[t]
\centering
% \caption{Examples of adversarial code generated by different methods.}
% \label{tab:code_comparison}
\begin{tabular}{c p{0.9\textwidth}}
\hline
\textbf{Original Code} & \lstset{
 columns=fixed,       
 % numbers=left,                                        % 在左侧显示行号
 % numberstyle=\tiny\color{gray},                       % 设定行号格式
 frame=none,                                          % 不显示背景边框
 % backgroundcolor=\color[RGB]{245,245,244},            % 设定背景颜色
 keywordstyle=\color[RGB]{40,40,255},                 % 设定关键字颜色
 numberstyle=\footnotesize\color{darkgray},           
 commentstyle=\it\color[RGB]{0,96,96},                % 设置代码注释的格式
 stringstyle=\rmfamily\slshape\color[RGB]{128,0,0},   % 设置字符串格式
 showstringspaces=false,                              % 不显示字符串中的空格
 language=c++,                                        % 设置语言
}
\begin{lstlisting}
ssize_t seq_read_iter(struct kiocb *iocb, struct iov_iter *iter) {
    struct seq_file *m = file->private_data;
    size_t copied = 0;
    size_t n;
    // some codes
    while (1) {
        // some codes
    }
    return copied;
}
\end{lstlisting} \\
\hline
\textbf{ALERT} &  \lstset{
 columns=fixed,       
 % numbers=left,                                        % 在左侧显示行号
 % numberstyle=\tiny\color{gray},                       % 设定行号格式
 frame=none,                                          % 不显示背景边框
 % backgroundcolor=\color[RGB]{245,245,244},            % 设定背景颜色
 keywordstyle=\color[RGB]{40,40,255},                 % 设定关键字颜色
 numberstyle=\footnotesize\color{darkgray},           
 commentstyle=\it\color[RGB]{0,96,96},                % 设置代码注释的格式
 stringstyle=\rmfamily\slshape\color[RGB]{128,0,0},   % 设置字符串格式
 showstringspaces=false,                              % 不显示字符串中的空格
 language=c++,                                        % 设置语言
 moredelim=[is][\textcolor{red}]{@@}{@@},  % 自定义红色高亮
}
\begin{lstlisting}
ssize_t seq_read_iter(struct kiocb *iocb, struct iov_iter *iter) {
    struct seq_file *m = file->private_data;
    size_t @@read_bytes@@ = 0; 
    size_t @@chunk_size@@; 
   // some codes
    while (1) {
        // some codes
    }
    m->op->stop(m, p);
    m->count = 0;
    goto Done;
    return read_bytes;
}
\end{lstlisting}\\
\hline
\textbf{DIP} &  \lstset{
 columns=fixed,       
  % numbers=left,                                        % 在左侧显示行号
 % numberstyle=\tiny\color{gray},                       % 设定行号格式
 frame=none,                                          % 不显示背景边框
 % backgroundcolor=\color[RGB]{245,245,244},            % 设定背景颜色
 keywordstyle=\color[RGB]{40,40,255},                 % 设定关键字颜色
 numberstyle=\footnotesize\color{darkgray},           
 commentstyle=\it\color[RGB]{0,96,96},                % 设置代码注释的格式
 stringstyle=\rmfamily\slshape\color[RGB]{128,0,0},   % 设置字符串格式
 showstringspaces=false,                              % 不显示字符串中的空格
 language=c++,                                        % 设置语言
 moredelim=[is][\textcolor{red}]{@@}{@@},  % 自定义红色高亮
}
\begin{lstlisting}
ssize_t seq_read_iter(struct kiocb *iocb, struct iov_iter *iter) {
    struct seq_file *m = file->private_data;
    size_t copied = 0;
    size_t n;
    // some codes
    @@string m_2 @@= "if (unlikely(err)) m->count = 0;";
    while (1) {
        // some codes
    }
    return copied; 
}
\end{lstlisting}\\
\hline
\textbf{HogVul} &  \lstset{
 columns=fixed,       
  % numbers=left,                                        % 在左侧显示行号
 % numberstyle=\tiny\color{gray},                       % 设定行号格式
 frame=none,                                          % 不显示背景边框
 % backgroundcolor=\color[RGB]{245,245,244},            % 设定背景颜色
 keywordstyle=\color[RGB]{40,40,255},                 % 设定关键字颜色
 numberstyle=\footnotesize\color{darkgray},           
 commentstyle=\it\color[RGB]{0,96,96},                % 设置代码注释的格式
 stringstyle=\rmfamily\slshape\color[RGB]{128,0,0},   % 设置字符串格式
 showstringspaces=false,                              % 不显示字符串中的空格
 language=c++,                                        % 设置语言
 moredelim=[is][\textcolor{red}]{@@}{@@},  % 自定义红色高亮
}
\begin{lstlisting}
ssize_t seq_read_iter(struct kiocb *iocb, struct iov_iter *iter) {
    struct seq_file *m = file->private_data;
    size_t @@total_read@@ = 0;
    size_t n;
     // some codes
    @@for (;;) {@@
        // some codes
    }
    return total_read;
}
\end{lstlisting}\\
\hline
\end{tabular}
\caption{Examples of adversarial code generated by different methods.}
\label{tab:code_comparison}
\end{table*}

\subsection*{G Comparison With More Advanced Baselines}
\label{sec:app_baselines}
To further underscore the novelty and superiority of HogVul, we contextualize its performance against contemporaneous state-of-the-art adversarial attack methods, we chose ITGen~\cite{ITGen}, SLODA~\cite{SLODA}, CODA~\cite{coda} and GraphCodeAttack~\cite{nguyen2024adversarialattackscodemodels} as more powerful baselines based on their outstanding performance. Results are presented in Table~\ref{tab:broader_baselines}.

On the Devign dataset, HogVul consistently outperforms all baselines across victim models. Notably, it achieves the highest ASR on GraphCodeBERT~(\textbf{91.60\%}) and CodeT5~(\textbf{97.28\%}), surpassing the runner-up by \textbf{2.49\%} and \textbf{1.03\%}, respectively. Moreover, HogVul attains an average ASR of \textbf{92.22\%}, demonstrating its highly competitive in comparison with high-quality methods.

\subsection*{H Comparison under the same budget}
\label{sec:same_budget}
For a more equitable assessment of HogVul’s efficiency, we performed supplementary experiments under identical query budgets. On the Devign dataset, we tallied the number of attacks and the corresponding attack success rates for the original baseline under 5,000, 10,000, and 30,000 queries respectively. The results are shown in the Table~\ref{tab:budget}. We respectively tested the number of query samples under different budgets, the attack success rate and counted the number of successful queries in a single attempt. As indicated, HogVul access the highest total query samples across all budget settings and achieves a significantly higher ASR than both baselines. This underscores HogVul’s superior efficiency-effectiveness trade-off for vulnerability attack tasks under constrained query budgets.

\subsection*{I Qualitative Example}
\label{sec:example}
In this section, we present the adversarial examples generated by our method, HogVul, alongside those produced by baseline methods. As shown in Table~\ref{tab:code_comparison}, the limitations of baseline approaches become evident upon analysis. ALERT generates adversarial examples by modifying variable names. While this approach is simple, it has limited practical impact because models are often insensitive to changes in identifiers, and such perturbations can be easily detected using static analysis tools~\cite{mai2020simultaneously,7765073}. On the other hand, DIP generates adversarial examples by inserting dead code that can be compiled~\cite{pang2018towards}. However, the inserted code is unrelated to the original logic, poorly concealed, and easily flagged as suspicious by developers or automated defense mechanisms.
In contrast, HogVul employs a more sophisticated approach by combining structural transformations (e.g., converting a while loop into a for loop) with variable substitution. This dual strategy ensures that the generated adversarial code not only preserves syntactic naturalness but also disrupts the model’s ability to recognize code patterns by altering the code structure. As a result, HogVul produces adversarial examples that are significantly harder to detect and defend against, both manually and through automated tools.

\begin{table*}
  \centering
  \begin{tabular}{l l c c c}
    \toprule
    Metric                & Method  & Budget 5000 & Budget 10000 & Budget 30000 \\
    \midrule
    \multirow{3}{*}{Total Number of Query Samples\textbf{↑}} 
                          & ALERT   & 21          & 30           & 76           \\
                          & DIP     & 25          & 41           & 135         \\
                          & HogVul  & \textbf{69}          & \textbf{142}          & \textbf{436}          \\
    \midrule
    \multirow{3}{*}{Attack Success Rate (\%) \textbf{↑}} 
                          & ALERT   & 62.85       & 63.33        & 66.05        \\
                          & DIP     & 58.00       & 63.65        & 70.74        \\
                          & HogVul  & \textbf{85.65}       & \textbf{85.36}        & \textbf{88.18}        \\
    \midrule
    Average Single Successful Query Count\textbf{↓} 
                          & ALERT   & \multicolumn{3}{c}{600} \\
                          & DIP     & \multicolumn{3}{c}{463} \\
                          & HogVul  & \multicolumn{3}{c}{\textbf{124}}  \\
    \bottomrule
  \end{tabular}
  \caption{Performance comparison of ALERT, DIP, and HogVul on the Devign dataset under fixed query budgets.}
  \label{tab:budget}
\end{table*}

\end{document}